\documentclass[a4paper, 11pt]{article}
\usepackage[dvipsnames]{xcolor}
\usepackage{url}

\usepackage{comment}
\usepackage{amsmath}
\usepackage{mathtools}
\usepackage{amssymb}
\usepackage{amsthm}
\usepackage{setspace}
\usepackage{placeins}
\usepackage{minibox}
\usepackage{graphicx}
\usepackage{subcaption}
\usepackage{tikz}
\usepackage[round]{natbib}
\usepackage{caption}
\usepackage{booktabs}
\usepackage{multirow}
\usepackage{authblk}
\usepackage{circledsteps}
\usepackage{fancyhdr}
\usepackage{lastpage}
\usepackage{hyperref}
\usepackage{cleveref}
\usepackage{autonum}
\usepackage{tabularx}
\usepackage[edges]{forest}
\usepackage[hmargin = 2cm, vmargin = 2cm]{geometry}
\usepackage{color}

\newcommand{\qf}[1]{F_{#1}^{-1}}

\newcommand{\Id}{\operatorname{Id}}

\newcommand{\push}[2]{#1_{\#}#2}

\newcommand{\otmap}[2]{T_{#1\to #2}}

\newcommand{\Bmap}{\mathcal B}
\newcommand{\Smap}{\mathcal S}
\newcommand{\Emap}{\mathcal E}

\newcommand{\W}{\mathbb W_2}
\newcommand{\R}{\mathbb{R}}
\newcommand{\F}{\mathcal F}

\newcommand{\E}{\mathbb E}
\newcommand{\Var}{\mathop{\rm Var}}

\DeclareMathOperator*{\argmin}{arg\,min}

\newtheorem{theorem}{Theorem}
\newtheorem{proposition}{Proposition}

\newtheorem{definition}{Definition}
\theoremstyle{definition}

\title{Distributional Holt--Winters}
\author{Peiwen Jiang\qquad Wilson Ye Chen\qquad Takuo Matsubara\qquad Minh-Ngoc Tran}
\affil{The University of Sydney Business School, Australia}
\date{}
\hypersetup{
  pdftitle={Distributional Holt--Winters},
  pdfauthor={Peiwen Jiang; Wilson Ye Chen; Takuo Matsubara; Minh-Ngoc Tran},
  pdfsubject={Distribution-valued time series forecasting},
  pdfkeywords={Distribution-valued time series; exponential smoothing; trend and seasonality; optimal transport; Wasserstein geometry}
}

\begin{document}
\maketitle

\begin{abstract}
Forecasting time series of probability distributions requires methods that accommodate changes in both location and shape while preserving valid distributions. We develop distributional Holt--Winters (DHW), a recursive method for time series of univariate distributions with level, trend and seasonality. The level is a probability distribution, and trend and seasonal effects are represented by monotone transport maps. This construction extends Wasserstein exponential smoothing and recovers classical additive Holt--Winters smoothing in the point-mass translation case. We give conditions under which the updates preserve valid distributions, and derive an exact transport-based error-correction representation. Parameters are estimated by minimising one-step squared Wasserstein prediction loss. Under conditionally centred innovations, we establish one-step conditional optimality and an excess-risk decomposition. A separate finite-horizon result identifies the smoothing parameters conditional on a common pre-update state, under regularity and non-degeneracy conditions. Simulations isolate the contributions of trend and adaptive seasonality, showing the value of updating seasonal maps when seasonality evolves. Applications to subway ridership, electricity use and retail turnover yield lower mean one-step loss than the considered distributional and seasonal benchmarks. Fitted component maps illustrate how trend and seasonal transformations can affect different parts of a distribution.
\end{abstract}



\section{Introduction}
\label{sec:introduction}

Many modern time series are naturally observed as sequences of probability distributions rather than scalar or vector observations. Examples arise when each time point contains a cross-sectional sample, a set of high-frequency observations, or a population of heterogeneous units. In finance, one may observe the distribution of intraday returns or cross-sectional asset returns over time; in economics and business analytics, one may observe evolving distributions of income, prices, demand, customer behaviour, or transaction values. In such settings, the forecasting target is the full distribution, from which summaries of location, dispersion and tail behaviour can be examined.

Distribution-valued time series require methods that respect the geometry and ordering structure of probability measures. Density functions are constrained to be nonnegative and integrate to one, while quantile functions must be nondecreasing. Standard linear operations on functions may therefore produce objects that no longer correspond to valid probability distributions. Wasserstein geometry provides a natural framework because it treats probability distributions as the primary statistical objects and compares them through optimal transport. This perspective has led to a growing literature on statistical modelling in Wasserstein space \citep{panaretos2019statistical}, including forecasting methods for density functions \citep{kokoszka2019forecasting}. Examples include Wasserstein autoregressive models (WAR) \citep{zhang2022wasserstein}, autoregressive transport models (ATM) \citep{atmzhu2023}, distributional autoregressive models (DAM) based on iterated transportation \citep{ghodrati2024distributional}, and distribution-on-distribution regression models (DoDRM) \citep{ghodrati2022distribution}. These four model families provide the distributional benchmarks considered in our empirical study.

A separate and highly influential tradition in forecasting is exponential smoothing. Classical exponential smoothing methods are widely used because they are simple, interpretable, and effective for business and economic time series. The Holt--Winters method extends simple exponential smoothing by decomposing a scalar time series into level, trend, and seasonal components, with each component updated recursively in response to forecast errors \citep{holt2004forecasting,winters1960forecasting}. This error-correction structure is central to the appeal of the method: the model is easy to estimate, easy to forecast with, and provides a transparent decomposition of the dynamics. State space formulations of exponential smoothing further place these recursions within a coherent statistical modelling framework \citep{hyndman2002state,hyndman2008forecasting}.

Recent work on Wasserstein exponential smoothing (WES) extends simple exponential smoothing from scalar observations to distribution-valued observations by replacing Euclidean interpolation with interpolation along Wasserstein geodesics \citep{matsubara2026WES}. The scalar forecast error is replaced by an optimal transport from the forecast distribution to the observed distribution, and the forecast is updated by moving part of the way along this transport. This provides a natural local-level model for distribution-valued time series. However, simple exponential smoothing is not designed to represent trend or seasonality. This is a substantive limitation in applications where distributions exhibit persistent directional movement or recurring seasonal changes in location, dispersion, or shape.

This paper develops distributional Holt--Winters (DHW) for time series of univariate distributions. In the scalar method, trend and seasonality act through additive shifts. DHW replaces these shifts by order-preserving transformations of an entire distribution. A translation map shifts all quantiles by the same amount, whereas a nonlinear monotone map can change dispersion or act more strongly in one tail. A level distribution is first transformed by a trend map and then by a seasonal map to produce the one-step-ahead forecast.

After a new distribution is observed, DHW removes the seasonal transformation and updates the level by Wasserstein interpolation. The trend and seasonal maps are updated by averaging their previous states with transports implied by the observation. These recursions share a transport-based forecast innovation, preserving the error-correction interpretation of classical smoothing. Under the stated invertibility, closure and moment conditions on the map class, the resulting distributions remain valid with finite second moments, without a separate projection at each update. The effects of fitted trend and seasonal maps can be examined by evaluating their displacements at selected quantiles.

The framework includes two familiar models as special cases. With point-mass distributions and translation maps, the recursion reduces algebraically to classical additive Holt--Winters smoothing. With identity trend and seasonal maps and their updates switched off, it reduces to WES. Multi-step forecasts are obtained by repeated composition of the fitted trend map and reuse of the latest seasonal maps.

The paper makes three main contributions. First, it introduces a monotone-map extension of Holt--Winters smoothing with an exact transport-based error-correction representation. Second, it establishes a conditional one-step Wasserstein risk decomposition and optimality result under conditionally centred innovations, together with finite-horizon identification of the smoothing parameters conditional on a common pre-update state and the stated regularity conditions. It also distinguishes these one-step results from general multi-step plug-in forecasting. Third, it develops estimation by one-step Wasserstein prediction loss and a quantile-based numerical implementation. A simulation study examines the roles of trend and adaptive seasonality, while three empirical applications evaluate forecast accuracy and illustrate the fitted distributional components.

The remainder of the paper is organised as follows. Section~\ref{sec:background} reviews classical additive Holt--Winters smoothing and Wasserstein exponential smoothing. Section~\ref{sec:dhw-transport-maps} introduces the monotone-map DHW recursion and gives conditions for preserving invertible maps and finite second moments. Section~\ref{sec:quantile-representation} develops its quantile and error-correction representations and the connections with classical Holt--Winters smoothing and WES. Section~\ref{sec:estimation-forecasting} discusses estimation and forecasting. Section~\ref{sec:theory} studies the forecasting properties of the model and identification of the smoothing parameters. Section~\ref{sec:simulation-study} reports a simulation study on finite-sample performance. Section~\ref{sec:empirical-applications} presents empirical applications, and Section~\ref{sec:conclusion} concludes.


\section{Background}
\label{sec:background}

\subsection{Classical Additive Holt--Winters}
\label{sec:classical-additive-holt-winters}

Classical exponential smoothing methods update a small number of latent components recursively as new observations arrive. The Holt--Winters method extends simple exponential smoothing by maintaining separate components for level, trend, and seasonality \citep{holt2004forecasting,winters1960forecasting}. We focus on the additive Holt--Winters specification, which provides the scalar benchmark for the distributional construction developed below.

Let $y_t\in\R$ be the observation at time $t$, and let $\ell_t$, $b_t$, and $s_t$ denote the level, trend, and seasonal components, respectively. Let $m$ be the seasonal period, measured in observations per cycle. Given the state at time $t-1$, the one-step-ahead forecast is
\begin{equation}
\hat y_t
=
\ell_{t-1}+b_{t-1}+s_{t-m}.
\label{eq:classical-hw-forecast}
\end{equation}
We use the following additive Holt--Winters updates:
\begin{align}
\ell_t
&=
(1-\alpha)(\ell_{t-1}+b_{t-1})
+
\alpha(y_t-s_{t-m}),
\label{eq:classical-hw-level}
\\
b_t
&=
(1-\beta)b_{t-1}
+
\beta(\ell_t-\ell_{t-1}),
\label{eq:classical-hw-trend}
\\
s_t
&=
(1-\gamma)s_{t-m}
+
\gamma(y_t-\ell_{t-1}-b_{t-1}),
\label{eq:classical-hw-season}
\end{align}
where $\alpha,\beta,\gamma\in[0,1]$ are smoothing parameters. The seasonal coefficient $\gamma$ directly controls forecast-error correction. Writing the seasonal update instead as $s_t=(1-\gamma^*)s_{t-m}+\gamma^*(y_t-\ell_t)$ gives $\gamma=(1-\alpha)\gamma^*$, so $\gamma^*\in[0,1]$ corresponds to $0\leq\gamma\leq1-\alpha$. We use the direct coefficient range $\gamma\in[0,1]$ in the formulation above.

The three equations have closely related interpretations. The level update combines the trend-adjusted previous level,
\[
\ell_{t-1}+b_{t-1},
\]
with the deseasonalised observation,
\[
y_t-s_{t-m}.
\]
The trend update averages the previous trend $b_{t-1}$ with the latest level increment $\ell_t-\ell_{t-1}$. The seasonal update averages the previous effect for the same season $s_{t-m}$ with the latest deviation of the observation from the trend-adjusted level,
\[
y_t-\ell_{t-1}-b_{t-1}.
\]
These three operations provide the template for the distributional model developed in Section~\ref{sec:dhw-transport-maps}: subtraction of a scalar seasonal effect will be replaced by inversion of a seasonal transport map, while scalar increments will be replaced by optimal transport maps between probability distributions.

The recursion can equivalently be expressed through the one-step-ahead forecast error
\begin{equation}
e_t
=
y_t-\hat y_t.
\label{eq:classical-hw-error}
\end{equation}
Using \eqref{eq:classical-hw-forecast}, the level and seasonal updates become
\begin{align}
\ell_t
&=
\ell_{t-1}+b_{t-1}+\alpha e_t,
\label{eq:classical-hw-level-error}
\\
s_t
&=
s_{t-m}+\gamma e_t.
\label{eq:classical-hw-season-error}
\end{align}
Moreover,
\[
\ell_t-\ell_{t-1}
=
b_{t-1}+\alpha e_t,
\]
so the trend update can be written as
\begin{equation}
b_t
=
b_{t-1}+\alpha\beta e_t.
\label{eq:classical-hw-trend-error}
\end{equation}
Thus the component updates above and the error-correction representation are algebraically equivalent. The former will be particularly useful for motivating the monotone-map construction, while the latter highlights the common innovation driving the three state updates.

We centre the initial seasonal effects to specify the initial allocation between level and seasonality:
\begin{equation}
\sum_{j=0}^{m-1}s_{-j}=0,
\label{eq:classical-seasonal-normalisation}
\end{equation}
This convention is imposed only at initialisation. Adding a constant to the level and subtracting the same constant from every seasonal component leaves the forecast unchanged. Subsequent updates need not preserve the zero sum. The corresponding distributional convention replaces the zero-average initial seasonal effect by an average initial seasonal map equal to the identity.

The $h$-step-ahead forecast from time $t$ is
\begin{equation}
\hat y_{t+h\mid t}
=
\ell_t+h b_t+s_{j(t,h)},
\label{eq:classical-hw-multistep}
\end{equation}
where $j(t,h)$ denotes the most recently updated seasonal index corresponding to time $t+h$. In the distributional model, additive trend extrapolation will be replaced by repeated composition of the estimated trend map.

State space formulations of exponential smoothing provide a statistical interpretation of these recursions and their forecasting properties; see, for example, \citet{hyndman2002state} and \citet{hyndman2008forecasting}.

\subsection{One-Dimensional Wasserstein Transport and Exponential Smoothing}
\label{sec:wes-transport-residuals}

We next review the one-dimensional Wasserstein constructions used throughout the paper. Let $\W(\R)$ denote the space of probability measures on $\R$ with finite second moment, equipped with the quadratic Wasserstein distance $W_2$. For $\rho\in\W(\R)$, let $\qf{\rho}$ denote its quantile function. For $\rho,\eta\in\W(\R)$,
\begin{equation}
W_2^2(\rho,\eta)
=
\int_0^1
\left\{
\qf{\rho}(u)-\qf{\eta}(u)
\right\}^2
\,du.
\label{eq:w2-quantile}
\end{equation}
Thus the Wasserstein distance in one dimension compares distributions through their corresponding quantiles.

Optimal transport also has a particularly simple representation in one dimension. When $\rho$ is atomless, the optimal transport map from $\rho$ to $\eta$ is uniquely determined $\rho$-almost everywhere and has the representation
\begin{equation}
\otmap{\rho}{\eta}
=
\qf{\eta}\circ F_\rho,
\label{eq:one-dimensional-ot}
\end{equation}
and satisfies
\begin{equation}
\push{(\otmap{\rho}{\eta})}{\rho}
=
\eta.
\label{eq:ot-pushforward}
\end{equation}
Equivalently,
\begin{equation}
\otmap{\rho}{\eta}\{\qf{\rho}(u)\}
=
\qf{\eta}(u),
\qquad\text{for almost every }u\in(0,1).
\label{eq:ot-quantile-matching}
\end{equation}

More generally, let $T:\R\to\R$ be a continuous nondecreasing map for which the relevant pushforward has finite second moment. If
\[
\eta=\push{T}{\rho},
\]
then the quantile function of $\eta$ is
\begin{equation}
\qf{\eta}
=
T\circ\qf{\rho}.
\label{eq:monotone-map-quantile}
\end{equation}
This identity will be central to our treatment of distributional trend and seasonality. Rather than adding unrestricted functions to a quantile function, we will transform a distribution by applying nondecreasing maps to its values.

Nondecreasing maps have several closure properties that are particularly useful for recursive forecasting. Their compositions are nondecreasing. If $T$ is a strictly increasing continuous bijection of $\R$, then $T^{-1}$ is also a strictly increasing continuous bijection. Finally, if $T_1$ and $T_2$ are nondecreasing, then for any $\omega\in[0,1]$,
\begin{equation}
(1-\omega)T_1+\omega T_2
\label{eq:convex-monotone-map}
\end{equation}
is nondecreasing. When both maps are strictly increasing continuous bijections, their composition and every such convex combination are also strictly increasing continuous bijections. These properties preserve ordering and invertibility; preservation of finite second moments requires separate conditions.

A related construction is Wasserstein interpolation. Given $\rho,\eta\in\W(\R)$ and $\omega\in[0,1]$, and assuming that the optimal transport map from $\rho$ to $\eta$ exists, the displacement interpolation from $\rho$ towards $\eta$ is
\begin{equation}
\rho_\omega
=
\push{
\left\{
(1-\omega)\Id
+
\omega\otmap{\rho}{\eta}
\right\}
}{\rho}.
\label{eq:wasserstein-interpolation}
\end{equation}
By \eqref{eq:ot-quantile-matching}, its quantile function is
\begin{equation}
\qf{\rho_\omega}(u)
=
(1-\omega)\qf{\rho}(u)
+
\omega\qf{\eta}(u),
\qquad u\in(0,1).
\label{eq:wasserstein-interpolation-quantile}
\end{equation}
The quantile formula also defines the interpolation for arbitrary $\rho,\eta\in\W(\R)$, since a convex combination of their nondecreasing, square-integrable quantile functions is again a quantile function in $\W(\R)$.

Wasserstein exponential smoothing uses this interpolation to extend simple exponential smoothing to distribution-valued observations \citep{matsubara2026WES}. Let $\nu_t\in\W(\R)$ denote the observed distribution at time $t$, and let $\mu_t\in\W(\R)$ denote its one-step-ahead predictor. After observing $\nu_t$, the next predictor is
\begin{equation}
\mu_{t+1}
=
\push{
\left\{
(1-\theta)\Id
+
\theta\otmap{\mu_t}{\nu_t}
\right\}
}{\mu_t},
\qquad
\theta\in[0,1].
\label{eq:wes-update}
\end{equation}
In quantile coordinates,
\begin{equation}
\qf{\mu_{t+1}}(u)
=
(1-\theta)\qf{\mu_t}(u)
+
\theta\qf{\nu_t}(u),
\qquad u\in(0,1).
\label{eq:wes-quantile-update}
\end{equation}
This is the Wasserstein analogue of scalar simple exponential smoothing: the current predictor is moved a fraction $\theta$ towards the latest observation. Importantly, the update remains within $\W(\R)$ by construction.

Wasserstein exponential smoothing is a local-level model and does not separately represent trend or seasonality. Extending it to Holt--Winters dynamics requires analogues of the three operations in \eqref{eq:classical-hw-level}--\eqref{eq:classical-hw-season}: trend adjustment, deseasonalisation, and recursive updating of trend and seasonal effects. The monotone-map formulation introduced in the next section provides these analogues directly. Trend adjustment is performed by pushing the level distribution through a strictly increasing trend map; seasonality is added and removed through a strictly increasing seasonal map and its inverse; and the trend and seasonal states are updated through convex combinations of optimal transport maps. These operations preserve monotonicity by construction and provide a direct transport-map counterpart to classical additive Holt--Winters smoothing.


\section{Distributional Holt--Winters via Monotone Transport Maps}
\label{sec:dhw-transport-maps}

We now introduce the distributional Holt--Winters model. The central idea is to represent trend and seasonality directly through monotone transport maps. The recursion preserves ordering through composition, inversion and convex averaging, with the following conditions ensuring that the resulting distributions remain in $\W(\R)$.

Let $\mathcal M$ be a class of strictly increasing continuous bijections $T:\R\to\R$ containing $\Id$. We assume that $\mathcal M$ is closed under composition, inversion and pointwise convex combinations, that its maps preserve finite second moments for the distributions to which they are applied below, and that the optimal transports used in the recursion admit representatives in $\mathcal M$. The moment condition applies also to inverse maps, since $\mathcal M$ is closed under inversion.

A sufficient example is the class of strictly increasing continuous bijections with bounded displacement, $\sup_x|T(x)-x|<\infty$. Inversion preserves the displacement bound, composition adds the bounds, and convex averaging preserves them. The bound $|T(x)|\leq |x|+C_T$ then ensures preservation of finite second moments. The numerical maps with unit-slope tails described in Section~\ref{sec:dhw-numerical-implementation} belong to this class.

Transport maps are determined only up to sets of source measure zero. When used as state maps, their representatives in $\mathcal M$ specify their values on all of $\R$, including outside the source support. A continuous strictly increasing transport on a compact interval support can be extended to a bijection of $\R$ using unit-slope tails. Such continuation does not remove flat segments within the support; the strictly increasing numerical representation is described in Section~\ref{sec:dhw-numerical-implementation}.

Let $\nu_t\in\W(\R)$ denote the observed distribution at time $t$, and let $m$ be the seasonal period, measured in observations per cycle. At time $t-1$, the filtered state consists of the level distribution $\lambda_{t-1}\in\W(\R)$, the trend map $\Bmap_{t-1}\in\mathcal M$, and the $m$ most recent seasonal maps $\Smap_{t-m},\ldots,\Smap_{t-1}\in\mathcal M$. Seasonal positions recur modulo $m$, while the map for each position is updated as new observations arrive. The one-step-ahead forecast distribution for $\nu_t$ is denoted by $\mu_t$.

Given the filtered state at time $t-1$, the trend-adjusted level distribution is obtained by pushing the previous level through the current trend map,
\begin{equation}
\eta_t
=
\push{(\Bmap_{t-1})}{\lambda_{t-1}}.
\label{eq:map-eta}
\end{equation}
The seasonal map associated with the current season is then applied to obtain the one-step-ahead forecast,
\begin{equation}
\mu_t
=
\push{(\Smap_{t-m})}{\eta_t}
=
\push{(\Smap_{t-m}\circ\Bmap_{t-1})}{\lambda_{t-1}}.
\label{eq:map-mu}
\end{equation}
Thus the distributional forecast has the same basic decomposition as classical additive Holt--Winters: the previous level is first adjusted for trend and then for seasonality. The distinction is that both adjustments are now transformations of probability distributions rather than scalar additions.

To complete the stochastic specification, let $\Emap_t\in\mathcal M$ denote a random innovation map. The observed distribution is generated from the one-step-ahead forecast according to
\begin{equation}
\nu_t
=
\push{(\Emap_t)}{\mu_t}.
\label{eq:dhw-observation}
\end{equation}
Thus the distributional Holt--Winters model has the sequential representation
\[
\lambda_{t-1}
\xrightarrow{\;\Bmap_{t-1}\;}
\eta_t
\xrightarrow{\;\Smap_{t-m}\;}
\mu_t
\xrightarrow{\;\Emap_t\;}
\nu_t.
\]
Conditional on an initial state and a sequence of innovation maps, this construction generates the distribution-valued time series recursively. We leave the distribution of $\Emap_t$ otherwise unspecified at this stage. In Section~\ref{sec:theory}, we introduce a conditional centring condition under which $\mu_t$ is the conditional Fr\'echet mean of $\nu_t$ under squared Wasserstein loss, and use this property to study identification of the smoothing parameters.

Since $\Emap_t$ is strictly increasing and $\nu_t=\push{(\Emap_t)}{\mu_t}$, it is an optimal transport from $\mu_t$ to $\nu_t$. When $\mu_t$ is atomless, this transport is uniquely determined $\mu_t$-almost everywhere, and
\[
\Emap_t=\otmap{\mu_t}{\nu_t}
\qquad \mu_t\text{-almost everywhere}.
\]

After observing $\nu_t$, the current seasonal effect is removed by applying the inverse seasonal map,
\begin{equation}
\widetilde\nu_t
=
\push{(\Smap_{t-m}^{-1})}{\nu_t}.
\label{eq:deseasonalised-observation}
\end{equation}
This is the transport-map analogue of the scalar deseasonalised observation $y_t-s_{t-m}$ in Section~\ref{sec:classical-additive-holt-winters}. The updated level is obtained by moving a fraction $\alpha$ along the Wasserstein geodesic from the trend-adjusted level $\eta_t$ towards the deseasonalised observation $\widetilde\nu_t$:
\begin{equation}
\lambda_t
=
\push{
\left\{
(1-\alpha)\Id
+
\alpha\otmap{\eta_t}{\widetilde\nu_t}
\right\}
}{\eta_t},
\qquad
\alpha\in[0,1].
\label{eq:map-level-update}
\end{equation}

The trend update follows the same logic as the scalar Holt--Winters update
\[
b_t
=
(1-\beta)b_{t-1}
+
\beta(\ell_t-\ell_{t-1}).
\]
The scalar level increment $\ell_t-\ell_{t-1}$ is replaced by the optimal transport map from the previous level distribution to the updated level distribution. Hence
\begin{equation}
\Bmap_t
=
(1-\beta)\Bmap_{t-1}
+
\beta\otmap{\lambda_{t-1}}{\lambda_t},
\qquad
\beta\in[0,1].
\label{eq:map-trend-update}
\end{equation}
Similarly, the scalar seasonal update averages the previous seasonal effect with the current deviation of the observation from the trend-adjusted level. Its distributional analogue is
\begin{equation}
\Smap_t
=
(1-\gamma)\Smap_{t-m}
+
\gamma\otmap{\eta_t}{\nu_t},
\qquad
\gamma\in[0,1].
\label{eq:map-season-update}
\end{equation}
All convex combinations of maps in \eqref{eq:map-level-update}--\eqref{eq:map-season-update} are understood pointwise.

\begin{definition}[Distributional Holt--Winters]
\label{def:dhw}
Given an initial state satisfying the map conditions above and $(\alpha,\beta,\gamma)\in[0,1]^3$, the DHW recursion consists of the forecast equations \eqref{eq:map-eta}--\eqref{eq:map-mu} and the level, trend and seasonal updates \eqref{eq:map-level-update}--\eqref{eq:map-season-update}. Together with the innovation maps and observation equation \eqref{eq:dhw-observation}, these equations define the distribution-valued process.
\end{definition}

The three smoothing parameters retain their familiar interpretation. The parameter $\alpha$ controls the responsiveness of the level distribution to the latest deseasonalised observation, $\beta$ controls the responsiveness of the trend map to the latest transport between level distributions, and $\gamma$ controls the responsiveness of the seasonal map to the latest seasonal transformation.

An initial centring convention is the map-space analogue of the zero-sum seasonal condition in \eqref{eq:classical-seasonal-normalisation}:
\begin{equation}
\frac{1}{m}
\sum_{j=0}^{m-1}
\Smap_{-j}(x)
=
x,
\qquad x\in\R.
\label{eq:map-seasonal-normalisation}
\end{equation}
Thus the average initial seasonal map is the identity map. When the seasonal maps are translations of the form $\Smap_j(x)=x+s_j$, condition \eqref{eq:map-seasonal-normalisation} reduces exactly to
\[
\sum_{j=0}^{m-1}s_{-j}=0.
\]
This condition centres the starting level--seasonality decomposition. It is imposed only at initialisation; the average seasonal map need not remain the identity as the maps are updated, and we do not recentre the seasonal states during filtering.

Under the stated map and moment conditions, the recursion preserves strictly increasing component maps and distributions with finite second moments throughout filtering.

\begin{proposition}[Monotonicity and Finite Second Moments]
\label{prop:map-admissibility}
Suppose that $\lambda_0\in\W(\R)$, the initial trend and seasonal maps belong to $\mathcal M$, the innovation maps $\Emap_t$ belong to $\mathcal M$, and the monotone optimal transport maps appearing in the recursion belong to $\mathcal M$. Then, for every $t$, the distributions $\eta_t$, $\mu_t$, $\nu_t$, $\widetilde\nu_t$, and $\lambda_t$ belong to $\W(\R)$, and the updated trend and seasonal maps remain in $\mathcal M$. In particular, the recursion does not require an additional monotonicity constraint or projection step.
\end{proposition}

\begin{proof}
Suppose that $\lambda_{t-1}\in\W(\R)$ and that the stored trend and seasonal maps belong to $\mathcal M$. Closure under composition and inversion, together with the assumed preservation of finite second moments, gives $\eta_t$, $\mu_t$, $\nu_t$ and $\widetilde\nu_t$ in $\W(\R)$. The level-correction map
\[
(1-\alpha)\Id
+
\alpha\otmap{\eta_t}{\widetilde\nu_t}
\]
belongs to $\mathcal M$ by closure under convex combinations, so \eqref{eq:map-level-update} gives $\lambda_t\in\W(\R)$. The trend and seasonal updates are also convex combinations of maps in $\mathcal M$ and therefore remain in that class. In particular, strict increase is preserved for all smoothing parameters in $[0,1]$, including their endpoints. Induction from the initial state completes the proof.
\end{proof}

Proposition~\ref{prop:map-admissibility} shows that, under its stated assumptions, no further monotonicity restriction on $(\alpha,\beta,\gamma)\in[0,1]^3$ is needed. Composition, inversion and convex averaging preserve the map structure, while the moment conditions ensure that the distributions remain in $\W(\R)$. The next section develops the quantile representation and its connection with classical Holt--Winters error correction.


\section{Quantile Representation}
\label{sec:quantile-representation}

The monotone-map recursion has a particularly transparent representation in quantile coordinates. For $\rho\in\W(\R)$, let $\qf{\rho}$ denote its quantile function. As reviewed in Section~\ref{sec:wes-transport-residuals}, if $T$ is continuous and nondecreasing, then
\[
\qf{\push{T}{\rho}}
=
T\circ\qf{\rho}.
\]
Applying this identity to \eqref{eq:map-eta}--\eqref{eq:deseasonalised-observation} gives
\begin{align}
\qf{\eta_t}
&=
\Bmap_{t-1}\circ\qf{\lambda_{t-1}},
\label{eq:quantile-map-eta}
\\
\qf{\mu_t}
&=
\Smap_{t-m}\circ\qf{\eta_t},
\label{eq:quantile-map-mu}
\\
\qf{\widetilde\nu_t}
&=
\Smap_{t-m}^{-1}\circ\qf{\nu_t}.
\label{eq:quantile-deseasonalised}
\end{align}
The level update \eqref{eq:map-level-update} therefore becomes
\begin{equation}
\qf{\lambda_t}(u)
=
(1-\alpha)\qf{\eta_t}(u)
+
\alpha
\Smap_{t-m}^{-1}
\left\{
\qf{\nu_t}(u)
\right\},
\qquad u\in(0,1).
\label{eq:quantile-map-level}
\end{equation}
Both $\qf{\eta_t}$ and $\Smap_{t-m}^{-1}\circ\qf{\nu_t}$ are nondecreasing functions of $u$, so their convex combination is again a valid quantile function.

The trend and seasonal updates also have simple quantile representations. Evaluating \eqref{eq:map-trend-update} at $\qf{\lambda_{t-1}}(u)$ gives
\begin{equation}
\Bmap_t
\left\{
\qf{\lambda_{t-1}}(u)
\right\}
=
(1-\beta)\qf{\eta_t}(u)
+
\beta\qf{\lambda_t}(u),
\label{eq:quantile-map-trend}
\end{equation}
while evaluating \eqref{eq:map-season-update} at $\qf{\eta_t}(u)$ gives
\begin{equation}
\Smap_t
\left\{
\qf{\eta_t}(u)
\right\}
=
(1-\gamma)\qf{\mu_t}(u)
+
\gamma\qf{\nu_t}(u).
\label{eq:quantile-map-season}
\end{equation}
Thus each update is an interpolation between two valid quantile functions, while the trend and seasonal effects are represented directly by monotone maps acting on values.

\subsection{Transport Error Correction}
\label{sec:transport-error-correction}

The recursion retains the error-correction interpretation of classical Holt--Winters smoothing. The key distinction is that the innovation can be represented before or after the seasonal transformation.

Define the forecast quantile residual by
\begin{equation}
R_t^{\mathrm{post}}(u)
=
\qf{\nu_t}(u)
-
\qf{\mu_t}(u),
\qquad u\in(0,1),
\label{eq:post-seasonal-residual}
\end{equation}
and the deseasonalised quantile residual by
\begin{equation}
R_t^{\mathrm{pre}}(u)
=
\Smap_{t-m}^{-1}
\left\{
\qf{\nu_t}(u)
\right\}
-
\qf{\eta_t}(u).
\label{eq:pre-seasonal-residual}
\end{equation}
The superscripts indicate whether the residual is expressed after or before the seasonal transformation. Both residuals represent the same underlying innovation map $\Emap_t$. The forecast quantile residual measures its effect directly in the forecast coordinates, whereas the deseasonalised quantile residual expresses it after removing the seasonal transformation with $\Smap_{t-m}^{-1}$. When the seasonal map is a translation, the two residuals coincide; for a nonlinear seasonal map, they generally differ.

Using \eqref{eq:quantile-map-level}, the level update can be written as
\begin{equation}
\qf{\lambda_t}
=
\qf{\eta_t}
+
\alpha R_t^{\mathrm{pre}}.
\label{eq:ec-level}
\end{equation}
Similarly, since
\[
\qf{\lambda_t}-\qf{\eta_t}
=
\alpha R_t^{\mathrm{pre}},
\]
equation \eqref{eq:quantile-map-trend} gives
\begin{equation}
\Bmap_t\circ\qf{\lambda_{t-1}}
=
\Bmap_{t-1}\circ\qf{\lambda_{t-1}}
+
\alpha\beta R_t^{\mathrm{pre}}.
\label{eq:ec-trend}
\end{equation}
Finally, \eqref{eq:quantile-map-season} gives
\begin{equation}
\Smap_t\circ\qf{\eta_t}
=
\Smap_{t-m}\circ\qf{\eta_t}
+
\gamma R_t^{\mathrm{post}}.
\label{eq:ec-season}
\end{equation}

Equations~\eqref{eq:ec-level}, \eqref{eq:ec-trend} and \eqref{eq:ec-season} provide the distributional analogue of the classical Holt--Winters error-correction form. The level and trend updates respond to the innovation after removing the current seasonal transformation, whereas the seasonal update responds to the innovation in the original forecast coordinates. All three updates are therefore driven by the same discrepancy between the forecast distribution and the observed distribution.

This relationship can also be seen directly in map space. The transport identities below hold almost everywhere with respect to their source measures. Under full support and continuity they hold globally; otherwise their global form uses compatible representatives outside the source support. By \eqref{eq:dhw-observation},
\[
\Emap_t
=
\otmap{\mu_t}{\nu_t}.
\]
Since
\[
\mu_t
=
\push{(\Smap_{t-m})}{\eta_t},
\]
the corresponding transports satisfy
\begin{align}
\otmap{\eta_t}{\nu_t}
&=
\Emap_t\circ\Smap_{t-m},
\label{eq:season-map-from-innovation}
\\
\otmap{\eta_t}{\widetilde\nu_t}
&=
\Smap_{t-m}^{-1}
\circ
\Emap_t
\circ
\Smap_{t-m}.
\label{eq:pullback-innovation}
\end{align}
Thus the innovation used to update the level is obtained by pulling $\Emap_t$ back through the current seasonal transformation.

For completeness, define the pre-seasonal and post-seasonal innovation displacements by
\[
\Delta_t^{\mathrm{pre}}
=
\Smap_{t-m}^{-1}
\circ
\Emap_t
\circ
\Smap_{t-m}
-
\Id,
\qquad
\Delta_t^{\mathrm{post}}
=
\Emap_t-\Id.
\]
The level-correction map
\begin{equation}
L_t
=
(1-\alpha)\Id
+
\alpha\otmap{\eta_t}{\widetilde\nu_t}
=
\Id+\alpha\Delta_t^{\mathrm{pre}}
\label{eq:level-correction-map}
\end{equation}
satisfies
\[
\lambda_t
=
\push{(L_t)}{\eta_t}.
\]
Moreover,
\[
\otmap{\lambda_{t-1}}{\lambda_t}
=
L_t\circ\Bmap_{t-1},
\]
so the trend and seasonal updates may be written as
\begin{align}
\Bmap_t
&=
\Bmap_{t-1}
+
\alpha\beta
\Delta_t^{\mathrm{pre}}
\circ
\Bmap_{t-1},
\label{eq:ec-map-trend}
\\
\Smap_t
&=
\Smap_{t-m}
+
\gamma
\Delta_t^{\mathrm{post}}
\circ
\Smap_{t-m}.
\label{eq:ec-map-season}
\end{align}
Equivalently, using $\Delta_t^{\mathrm{post}}=\Emap_t-\Id$, the seasonal update is
\[
\Smap_t
=
\Smap_{t-m}
+
\gamma
(\Emap_t-\Id)
\circ
\Smap_{t-m}.
\]
These expressions make explicit that the level, trend, and seasonal updates are all driven by the same underlying forecast innovation, represented in the coordinates appropriate to each component.

\subsection{Classical Holt--Winters and Wasserstein Exponential Smoothing}
\label{sec:benchmark-special-cases}

The proposed recursion contains classical additive Holt--Winters smoothing as a special case. Suppose all distributions are point masses,
\[
\lambda_t=\delta_{\ell_t},
\qquad
\nu_t=\delta_{y_t},
\]
and restrict the trend and seasonal maps to translations,
\begin{equation}
\Bmap_t(x)=x+b_t,
\qquad
\Smap_t(x)=x+s_t.
\label{eq:translation-maps}
\end{equation}
Then
\[
\eta_t
=
\delta_{\ell_{t-1}+b_{t-1}},
\qquad
\mu_t
=
\delta_{\ell_{t-1}+b_{t-1}+s_{t-m}},
\]
and
\[
\widetilde\nu_t
=
\delta_{y_t-s_{t-m}}.
\]
The level update therefore becomes
\begin{equation}
\ell_t
=
(1-\alpha)(\ell_{t-1}+b_{t-1})
+
\alpha(y_t-s_{t-m}).
\label{eq:classical-level-map}
\end{equation}

Choosing the translation representative of the transport from $\lambda_{t-1}$ to $\lambda_t$ gives
\[
\otmap{\lambda_{t-1}}{\lambda_t}(x)
=
x+\ell_t-\ell_{t-1},
\]
so the trend update becomes
\begin{equation}
b_t
=
(1-\beta)b_{t-1}
+
\beta(\ell_t-\ell_{t-1}).
\label{eq:classical-trend-map}
\end{equation}
Similarly,
\[
\otmap{\eta_t}{\nu_t}(x)
=
x+y_t-\ell_{t-1}-b_{t-1},
\]
which gives
\begin{equation}
s_t
=
(1-\gamma)s_{t-m}
+
\gamma(y_t-\ell_{t-1}-b_{t-1}).
\label{eq:classical-season-map}
\end{equation}
Equations~\eqref{eq:classical-level-map}--\eqref{eq:classical-season-map} recover the additive Holt--Winters recursion with the coefficient convention specified in Section~\ref{sec:classical-additive-holt-winters}.

Wasserstein exponential smoothing is obtained as the corresponding local-level special case. If the trend and seasonal maps are initialised as $\Id$ and $\beta=\gamma=0$, then they remain equal to $\Id$. Hence $\eta_t=\mu_t=\lambda_{t-1}$ and $\widetilde\nu_t=\nu_t$, so
\begin{equation}
\qf{\lambda_t}
=
(1-\alpha)\qf{\lambda_{t-1}}
+
\alpha\qf{\nu_t}.
\label{eq:dhw-wes-special-case}
\end{equation}
This is precisely the Wasserstein exponential smoothing update reviewed in Section~\ref{sec:wes-transport-residuals}.

\subsection{Interpreting Distributional Components}
\label{sec:map-interpretation}

The monotone-map formulation represents trend and seasonality as transformations of values. Their effects can nevertheless be examined at any probability level through a reference quantile function. For example, the displacement of the fitted trend map at rank $u$ of the current level distribution is
\begin{equation}
\Delta B_t(u)
=
\Bmap_t\{\qf{\lambda_t}(u)\}
-
\qf{\lambda_t}(u),
\label{eq:effective-trend-displacement}
\end{equation}
while the displacement of the updated seasonal map at rank $u$ of the trend-adjusted distribution is
\begin{equation}
\Delta S_t(u)
=
\Smap_t\{\qf{\eta_t}(u)\}
-
\qf{\eta_t}(u).
\label{eq:effective-seasonal-displacement}
\end{equation}
A translation map produces the same displacement at every probability level and therefore changes location without changing shape. A nonlinear monotone map can produce different displacements across the distribution. For example, a seasonal map may move the lower tail substantially while leaving the centre almost unchanged, or expand both tails relative to the centre. The map formulation therefore retains the ability to describe distribution-specific trend and seasonality while ensuring that the transformed distributions remain valid.

Figure~\ref{fig:retail-anatomy} illustrates these components for Australian retail turnover. To compare seasonal maps across phases and filtered states, it is useful to evaluate them at fixed reference quantiles. Appendix~\ref{app:empirical-components} defines the common reference values used in the empirical figures.


\section{Estimation and Forecasting}
\label{sec:estimation-forecasting}

We estimate the smoothing parameters by minimising one-step-ahead squared Wasserstein prediction loss. For a given parameter vector, the monotone-map recursion is applied sequentially to the observed distributions, producing a corresponding sequence of one-step-ahead forecasts. Multi-step forecasts are subsequently obtained by propagating the estimated state forward using repeated composition of the trend map and the most recently updated seasonal maps.

\subsection{Initialisation}
\label{sec:dhw-initialisation}

Let $\nu_1,\ldots,\nu_N$ denote the observed distribution-valued time series, supplied as distributions, samples or empirical quantile functions. The numerical representation is described in Section~\ref{sec:dhw-numerical-implementation}. The recursion requires an initial level distribution $\lambda_0$, an initial trend map $\Bmap_0$, and $m$ initial seasonal maps $\Smap_0,\ldots,\Smap_{1-m}$. These are chosen to satisfy the map and moment conditions of Section~\ref{sec:dhw-transport-maps} and the initial seasonal centring convention in \eqref{eq:map-seasonal-normalisation}.

The initial state may be specified directly or estimated from an initial segment of the observations. For example, suppose the initialisation window contains several complete seasonal cycles. For each seasonal position, let $\bar\nu_{-j}$ denote the one-dimensional Wasserstein barycentre of the observations corresponding to that position, where $j=0,\ldots,m-1$. The initial level distribution can then be defined by
\[
\qf{\lambda_0}(u)
=
\frac{1}{m}
\sum_{j=0}^{m-1}
\qf{\bar\nu_{-j}}(u),
\]
with initial seasonal maps
\[
\Smap_{-j}
=
\otmap{\lambda_0}{\bar\nu_{-j}}.
\]
This construction satisfies the initial seasonal centring condition on the support of $\lambda_0$. For strictly increasing numerical quantile ordinates, both the phase averages and their barycentre are strictly increasing, so the initial seasonal transports are strictly increasing. Since all these maps share the same source knots, the linear displacement interpolation and unit-slope tail continuation in Section~\ref{sec:dhw-numerical-implementation} preserve their average as $\Id$ on $\R$, up to rounding. The initial trend map may be set to the identity or estimated separately from the initial observations.

We treat the initial state as fixed when estimating the smoothing parameters, so that all candidate parameter values are evaluated from the same starting point. An initial burn-in period may be used to reduce the influence of initialisation on the prediction criterion.

\subsection{Parameter Estimation}
\label{sec:dhw-parameter-estimation}

Let $\boldsymbol{\theta}=(\alpha,\beta,\gamma)\in\Theta$, where $\Theta\subseteq[0,1]^3$ is the parameter space. For each candidate $\boldsymbol{\theta}$, the recursion in \eqref{eq:map-eta}--\eqref{eq:map-season-update} is applied to the observations, starting from the chosen initial state. Write $\mu_t(\boldsymbol{\theta})$ for the resulting one-step-ahead forecast distribution at time $t$. In particular, $\mu_t(\boldsymbol{\theta})$ is constructed using only observations available before time $t$.

Let $t_0$ denote the first time point included in the estimation criterion, following any initialisation or burn-in period. We define the estimator by
\begin{equation}
\widehat{\boldsymbol{\theta}}_N
\in
\argmin_{\boldsymbol{\theta}\in\Theta}
L_N(\boldsymbol{\theta}),
\qquad
L_N(\boldsymbol{\theta})
=
\frac{1}{N-t_0+1}
\sum_{t=t_0}^{N}
W_2^2\left\{
\nu_t,\mu_t(\boldsymbol{\theta})
\right\}.
\label{eq:dhw-estimation-objective}
\end{equation}
By the one-dimensional Wasserstein representation in \eqref{eq:w2-quantile}, the objective can equivalently be expressed as
\begin{equation}
L_N(\boldsymbol{\theta})
=
\frac{1}{N-t_0+1}
\sum_{t=t_0}^{N}
\int_0^1
\left[
\qf{\nu_t}(u)
-
\qf{\mu_t(\boldsymbol{\theta})}(u)
\right]^2
\,du.
\label{eq:dhw-estimation-quantile}
\end{equation}
Thus estimation requires only the observed quantile functions and the one-step-ahead quantile forecasts generated by the recursion. The forecasting interpretation of this criterion is established in Section~\ref{sec:theory}.

The error-correction representation in Section~\ref{sec:transport-error-correction} also suggests a useful computational reparameterisation. In that representation, the level and trend updates depend on $\alpha$ and the product $\alpha\beta$, respectively. Defining
\[
\delta=\alpha\beta,
\]
we may equivalently optimise over $(\alpha,\delta,\gamma)$ subject to
\[
0<\alpha\leq1,
\qquad
0\leq\delta\leq\alpha,
\qquad
0\leq\gamma\leq1,
\]
and recover
\[
\beta=\frac{\delta}{\alpha}.
\]
This parameterisation separates the immediate level response from the correction incorporated into the trend map and can be convenient for numerical optimisation. In the error-correction representation, setting $\alpha=0$ removes the trend correction, so the updates no longer depend on $\beta$; the reparameterisation is therefore used with $\alpha>0$.

\subsection{Multi-step Forecasting}
\label{sec:dhw-multistep-forecasting}

After estimating $\boldsymbol{\theta}$, the recursion is applied through time $N$ to obtain the fitted state, consisting of the level distribution $\lambda_N$, the trend map $\Bmap_N$, and the most recently updated seasonal maps. For notational simplicity, we suppress their dependence on $\widehat{\boldsymbol{\theta}}_N$.

We distinguish the one-step-ahead forecast $\mu_t$, formed before observing $\nu_t$, from a general $h$-step-ahead forecast $\widehat\mu_{N+h\mid N}$ formed using information available at time $N$. In particular, the fitted one-step forecast satisfies
\[
\widehat\mu_{N+1\mid N}
=
\mu_{N+1}(\widehat{\boldsymbol{\theta}}_N).
\]

For a forecast horizon $h\geq1$, let
\begin{equation}
j(N,h)
=
N+h-m\left\lceil\frac{h}{m}\right\rceil
\label{eq:dhw-seasonal-forecast-index}
\end{equation}
denote the index of the most recently updated seasonal map corresponding to the target time $N+h$. The $h$-step-ahead plug-in forecast is defined by
\begin{equation}
\widehat\mu_{N+h\mid N}
=
\push{
\left(
\Smap_{j(N,h)}
\circ
\Bmap_N^{\circ h}
\right)
}{\lambda_N},
\qquad h\geq1,
\label{eq:dhw-multistep-forecast}
\end{equation}
where $\Bmap_N^{\circ h}$ denotes the $h$-fold composition of the fitted trend map. Equivalently, its quantile function is
\begin{equation}
\qf{\widehat\mu_{N+h\mid N}}
=
\Smap_{j(N,h)}
\circ
\Bmap_N^{\circ h}
\circ
\qf{\lambda_N}.
\label{eq:dhw-multistep-quantile}
\end{equation}

The trend map is held fixed when extrapolating from the forecast origin, while the latest available seasonal maps are reused according to the seasonal period. In particular, for $1\leq h\leq m$,
\[
j(N,h)=N+h-m,
\]
so the first $m-1$ forecast horizons use previously stored seasonal maps, whereas the forecast at horizon $m$ uses the seasonal map $\Smap_N$ updated at the forecast origin.

For a translation trend map $\Bmap_N(x)=x+b_N$, repeated composition gives
\[
\Bmap_N^{\circ h}(x)=x+hb_N.
\]
Together with translation seasonal maps, \eqref{eq:dhw-multistep-forecast} therefore reduces to the classical additive Holt--Winters forecast in \eqref{eq:classical-hw-multistep}.

Since the forecast is constructed through compositions of increasing maps, it remains a valid distribution under the map and moment conditions imposed on the model. The construction is a deterministic plug-in forecast obtained by propagating the latest fitted state without future innovations. Under the conditions developed in Section~\ref{sec:theory}, the one-step forecast with the true smoothing parameters is optimal under squared Wasserstein loss, but this optimality does not generally extend to longer horizons.

\subsection{Numerical Implementation}
\label{sec:dhw-numerical-implementation}

The recursion and estimation criterion can be implemented on a common equally spaced probability grid
\[
0<u_1<\cdots<u_K<1.
\]
Writing the observed and forecast quantile functions on this grid, the one-step squared Wasserstein prediction loss is approximated by
\begin{equation}
L_{N,K}(\boldsymbol{\theta})
=
\frac{1}{(N-t_0+1)K}
\sum_{t=t_0}^{N}
\sum_{k=1}^{K}
\left[
\qf{\nu_t}(u_k)
-
\qf{\mu_t(\boldsymbol{\theta})}(u_k)
\right]^2.
\label{eq:dhw-estimation-grid}
\end{equation}
The grid criterion therefore approximates the squared Wasserstein loss by averaging squared differences between corresponding quantiles.

At each filtering step, the current quantile functions are evaluated using the compositions in \eqref{eq:quantile-map-eta}--\eqref{eq:quantile-deseasonalised}. The level is updated through \eqref{eq:quantile-map-level}, while the trend and seasonal maps are represented nonparametrically on a common value grid and updated by pointwise convex averaging. A monotone transport map specified by quantile pairs $x_k\mapsto y_k$ is evaluated by piecewise-linear interpolation of its displacement $y_k-x_k$. Outside the range of the stored knots, the boundary displacement is held constant, so that the completed map has unit slope in both tails. The same convention is used for evaluating stored maps, constructing transport maps, and evaluating inverse seasonal maps.

The numerical implementation requires the discretised quantile functions and the stored trend and seasonal maps to be strictly increasing. Empirical quantile functions containing repeated values are therefore preprocessed before constructing the corresponding transport interpolants; Appendix~\ref{app:empirical-numerics} describes the adjustments used in the applications. Strictly increasing source and target knots give strictly increasing interpolated maps, and the DHW convex updates preserve strictness for all smoothing parameters in $[0,1]$. No additional projection to restore quantile monotonicity is required during filtering.

Estimation uses the constrained reparameterisation in Section~\ref{sec:dhw-parameter-estimation}. The finite-dimensional objective is minimised over $(\alpha,\delta,\gamma)$, and $\beta=\delta/\alpha$ is recovered afterwards.

For out-of-sample rolling one-step-ahead forecasts, the estimated smoothing parameters are held fixed while the filtered state is updated sequentially as new observations become available. Each forecast is evaluated before the corresponding observation is incorporated into the recursion.

The DHW routines are implemented in the accompanying Python package
\texttt{dhwsmooth}, available at
\url{https://github.com/wilson-ye-chen/dhwsmooth}.
The package provides routines for filtering, estimation, forecasting,
simulation, and visualisation; basic usage is described in
Appendix~\ref{app:dhwsmooth}.


\section{Forecasting Properties}
\label{sec:theory}

This section establishes the forecasting properties of the proposed model. We first introduce conditions under which the one-step-ahead DHW forecast is optimal under squared Wasserstein loss and derive the corresponding risk decomposition. We then examine the distinction between plug-in multi-step forecasts and conditional Wasserstein-optimal forecasts. Finally, we establish a finite-horizon identification result that characterises how the three smoothing parameters affect the forecast response to a realised innovation.

\paragraph{Innovation conditions.}

Let $\F_t$ denote the information available after observing $\nu_t$, including the information used to initialise the recursion. In particular, the one-step-ahead forecast $\mu_t$ is $\F_{t-1}$-measurable. Recall from \eqref{eq:dhw-observation} that the observed distribution is generated according to
\[
\nu_t=\push{(\Emap_t)}{\mu_t}.
\]
Since $\Emap_t$ is increasing, its action on the quantile function satisfies
\[
\qf{\nu_t}(u)=\Emap_t\{\qf{\mu_t}(u)\},
\qquad u\in(0,1).
\]

We assume that the innovation map is conditionally centred along the forecast quantile function:
\begin{equation}
\E\left[
\Emap_t\{\qf{\mu_t}(u)\}-\qf{\mu_t}(u)
\mid\F_{t-1}
\right]=0,
\qquad\text{for almost every }u\in(0,1).
\label{eq:innovation-centering}
\end{equation}
We also assume that the innovation has finite expected squared Wasserstein magnitude,
\begin{equation}
\E\left[W_2^2(\nu_t,\mu_t)\right]<\infty.
\label{eq:innovation-second-moment}
\end{equation}
A sufficient formulation of \eqref{eq:innovation-centering}, subject to the usual conditional measurability and integrability requirements, is that the innovation map has conditional mean equal to the identity when evaluated at $\F_{t-1}$-measurable arguments. These conditions do not require independence or stationarity of the innovation maps, and their conditional distributions may vary over time.

Recall the forecast quantile residual $R_t^{\mathrm{post}}$ defined in \eqref{eq:post-seasonal-residual}. Under \eqref{eq:innovation-centering},
\begin{equation}
\E\left[R_t^{\mathrm{post}}(u)\mid\F_{t-1}\right]=0,
\qquad\text{for almost every }u\in(0,1).
\label{eq:post-seasonal-centering}
\end{equation}
Thus the forecast quantile residual is a martingale difference. This property need not extend to the deseasonalised quantile residual $R_t^{\mathrm{pre}}$, because removing seasonality involves composition with the potentially nonlinear inverse map $\Smap_{t-m}^{-1}$. This distinction will also be relevant when considering multi-step forecasting.

\paragraph{Wasserstein forecast risk.}

The conditional centring condition gives the one-step-ahead forecast $\mu_t$ a direct statistical interpretation. Let $\rho_t\in\W(\R)$ be any $\F_{t-1}$-measurable candidate forecast with finite expected squared Wasserstein loss. By the one-dimensional Wasserstein representation in \eqref{eq:w2-quantile},
\[
W_2^2(\nu_t,\rho_t)
=
\int_0^1
\left[
R_t^{\mathrm{post}}(u)
+
\qf{\mu_t}(u)
-
\qf{\rho_t}(u)
\right]^2
\,du.
\]
Taking conditional expectations and applying \eqref{eq:post-seasonal-centering} gives the decomposition
\begin{equation}
\E\left[
W_2^2(\nu_t,\rho_t)
\mid\F_{t-1}
\right]
=
\Omega_t
+
W_2^2(\mu_t,\rho_t),
\label{eq:wasserstein-risk-decomposition}
\end{equation}
where
\begin{equation}
\Omega_t
=
\E\left[
W_2^2(\nu_t,\mu_t)
\mid\F_{t-1}
\right]
\label{eq:irreducible-wasserstein-risk}
\end{equation}
is the irreducible one-step forecast risk.

Since the second term on the right-hand side of \eqref{eq:wasserstein-risk-decomposition} is nonnegative and vanishes only when $\rho_t=\mu_t$, the DHW forecast is the unique conditional Fr\'echet mean of $\nu_t$ under squared Wasserstein loss. The conditional centring assumption therefore identifies $\mu_t$ as the optimal one-step-ahead distributional forecast.

The irreducible risk has a particularly simple interpretation for translation innovations. If
\[
\Emap_t(x)=x+\xi_t,
\qquad
\E(\xi_t\mid\F_{t-1})=0,
\]
then
\[
W_2^2(\nu_t,\mu_t)=\xi_t^2,
\]
and consequently
\[
\Omega_t=\E(\xi_t^2\mid\F_{t-1}).
\]
For general innovation maps, $\Omega_t$ also captures random changes in dispersion and shape.

The risk decomposition provides the population justification for the one-step estimation criterion in \eqref{eq:dhw-estimation-objective}. Let $\boldsymbol{\theta}_0$ denote the data-generating smoothing parameters, and suppose that the model and its initial state are correctly specified, so that
\[
\mu_t=\mu_t(\boldsymbol{\theta}_0).
\]
For each fixed candidate parameter vector $\boldsymbol{\theta}\in\Theta$, the forecast $\mu_t(\boldsymbol{\theta})$ is constructed from the same initial state using observations available before time $t$. Applying \eqref{eq:wasserstein-risk-decomposition} gives
\begin{equation}
\E W_2^2\{\nu_t,\mu_t(\boldsymbol{\theta})\}
-
\E W_2^2\{\nu_t,\mu_t(\boldsymbol{\theta}_0)\}
=
\E W_2^2\{
\mu_t(\boldsymbol{\theta}_0),
\mu_t(\boldsymbol{\theta})
\}.
\label{eq:population-excess-risk}
\end{equation}

In particular, define the population counterpart of the estimation criterion by
\[
\mathcal R_N(\boldsymbol{\theta})
=
\frac{1}{N-t_0+1}
\sum_{t=t_0}^{N}
\E W_2^2\{\nu_t,\mu_t(\boldsymbol{\theta})\}.
\]
Provided the expectations are finite, \eqref{eq:population-excess-risk} implies
\begin{equation}
\mathcal R_N(\boldsymbol{\theta})
-
\mathcal R_N(\boldsymbol{\theta}_0)
=
\frac{1}{N-t_0+1}
\sum_{t=t_0}^{N}
\E W_2^2\{
\mu_t(\boldsymbol{\theta}_0),
\mu_t(\boldsymbol{\theta})
\}
\geq0.
\label{eq:population-average-excess-risk}
\end{equation}
Thus minimising expected one-step Wasserstein prediction loss targets the true conditional forecast. The equality identifies forecast distributions, although uniqueness of the smoothing parameters requires additional identification conditions. Moreover, the nonnegativity in \eqref{eq:population-average-excess-risk} concerns expected risk; a realised difference between sample prediction losses need not be nonnegative.

\paragraph{Multi-step forecasts.}

The optimality result above is intrinsically one-step ahead. For $h>1$, let $\mu^\star_{t+h\mid t}$ denote the conditional Fr\'echet mean of $\nu_{t+h}$ given $\F_t$. Applying the one-dimensional Wasserstein representation and iterated conditional expectations yields
\begin{equation}
\qf{\mu^\star_{t+h\mid t}}(u)
=
\E\left[
\qf{\nu_{t+h}}(u)
\mid\F_t
\right]
=
\E\left[
\qf{\mu_{t+h}}(u)
\mid\F_t
\right],
\label{eq:multistep-frechet-mean}
\end{equation}
for almost every $u\in(0,1)$, whenever the relevant conditional moments exist.

By contrast, the plug-in forecast in \eqref{eq:dhw-multistep-forecast} propagates the fitted state forward while holding the latest trend map fixed and reusing the most recently updated seasonal maps. Equivalently, it sets future innovations equal to the identity. Although this procedure is the direct distributional counterpart of classical Holt--Winters forecasting, the resulting forecast need not equal \eqref{eq:multistep-frechet-mean}. In particular, conditional expectation does not generally commute with nonlinear map composition.

To illustrate this distinction, consider the non-seasonal case with a conditionally centred translation innovation,
\[
\Emap_{t+1}(x)=x+\xi_{t+1},
\qquad
\E(\xi_{t+1}\mid\F_t)=0.
\]
For a fixed probability level $u$, write
\[
q=\qf{\lambda_t}(u),
\qquad
y=\Bmap_t(q).
\]
After observing the innovation at time $t+1$, the level and trend updates give
\[
\qf{\lambda_{t+1}}(u)=y+\alpha\xi_{t+1}
\]
and
\[
\Bmap_{t+1}(x)=\Bmap_t(x)+\alpha\beta\xi_{t+1}.
\]
Consequently, the one-step forecast formed at time $t+1$ satisfies
\begin{equation}
\qf{\mu_{t+2}}(u)
=
\Bmap_t\{y+\alpha\xi_{t+1}\}
+
\alpha\beta\xi_{t+1}.
\label{eq:two-step-random-forecast}
\end{equation}
Taking conditional expectations gives
\[
\qf{\mu^\star_{t+2\mid t}}(u)
=
\E\left[
\Bmap_t\{y+\alpha\xi_{t+1}\}
\mid\F_t
\right],
\]
whereas the plug-in two-step forecast has quantile function $\Bmap_t(y)$. These quantities coincide if $\Bmap_t$ is affine, but not generally when the trend map is nonlinear. For a twice continuously differentiable trend map and sufficiently small innovations, a second-order expansion gives the approximation
\[
\qf{\mu^\star_{t+2\mid t}}(u)
-
\qf{\mu_{t+2\mid t}}(u)
\approx
\frac{\alpha^2}{2}
\Var(\xi_{t+1}\mid\F_t)
\Bmap_t''(y),
\]
where $\mu_{t+2\mid t}$ denotes the plug-in forecast formed from the true state at time $t$. Thus nonlinear trend propagation can introduce a Jensen-type correction even when the innovation map is conditionally centred.

A discrepancy may also arise with affine maps when the innovation changes distributional shape. For example, suppose the seasonal maps are identities, $\Bmap_t=\Id$, and
\[
\Emap_{t+1}(x)=A_{t+1}x,
\qquad
A_{t+1}>0,
\qquad
\E(A_{t+1}\mid\F_t)=1.
\]
Writing $X_{t+1}=A_{t+1}-1$, the subsequent level and trend updates yield
\[
\qf{\mu_{t+2}}(u)
=
(1+\alpha X_{t+1})
(1+\alpha\beta X_{t+1})
\qf{\lambda_t}(u).
\]
It follows that
\[
\qf{\mu^\star_{t+2\mid t}}(u)
=
\left\{
1+\alpha^2\beta
\Var(A_{t+1}\mid\F_t)
\right\}
\qf{\lambda_t}(u),
\]
whereas the plug-in forecast is simply $\lambda_t$. In this case, the discrepancy arises from the interaction between the level and trend responses to the same random dilation.

There is nevertheless an important special case in which plug-in forecasting retains multi-step optimality. Suppose the trend and seasonal maps are affine and all future innovation maps are conditionally centred translations. Under the DHW updates, the map slopes then remain fixed, while the intercepts evolve linearly in the subsequent innovations. Iterated conditional expectations consequently eliminate the future innovation terms, and the plug-in forecast coincides with the conditional Fr\'echet mean at every horizon for which the relevant moments exist. Outside this affine-translation subclass, such an equivalence is not guaranteed. This distinction supports our use of one-step Wasserstein prediction loss as the estimation criterion, without requiring the plug-in multi-step forecasts to be conditionally optimal.

\paragraph{Identification of the smoothing parameters.}

The error-correction representation in Section~\ref{sec:transport-error-correction} reveals distinct dynamic roles for the smoothing parameters. The level responds directly to the pre-seasonal innovation through $\alpha$, the trend incorporates the same innovation through the product $\alpha\beta$, and the seasonal map responds to the post-seasonal innovation through $\gamma$. We now show that these effects can be separated through their finite-horizon forecast response.

Fix a time $t$, together with the pre-update state and the realised observation $\nu_t$. Write
\[
B=\Bmap_{t-1},
\qquad
S=\Smap_{t-m},
\qquad
\eta=\eta_t,
\qquad
\Emap=\Emap_t.
\]
Recalling the innovation displacements introduced in Section~\ref{sec:transport-error-correction}, we have
\[
\Delta_t^{\mathrm{pre}}
=
S^{-1}\circ \Emap\circ S-\Id,
\qquad
\Delta_t^{\mathrm{post}}
=
\Emap-\Id.
\]
The pre-update state and the observation are held fixed when comparing candidate smoothing parameters. In particular, $B$, $S$, $\eta$, $\Emap$, $\Delta_t^{\mathrm{pre}}$, and $\Delta_t^{\mathrm{post}}$ do not depend on the candidate parameter vector in the following argument.

For $\boldsymbol{\theta}=(\alpha,\beta,\gamma)$, introduce the maps
\begin{equation}
\begin{aligned}
L_\alpha&=\Id+\alpha\Delta_t^{\mathrm{pre}},\\
K_{\alpha,\beta}&=(\Id+\alpha\beta\Delta_t^{\mathrm{pre}})\circ B,\\
J_\gamma&=(\Id+\gamma\Delta_t^{\mathrm{post}})\circ S.
\end{aligned}
\label{eq:identification-maps}
\end{equation}
By \eqref{eq:level-correction-map}, \eqref{eq:ec-map-trend}, and \eqref{eq:ec-map-season}, these maps determine the updated state:
\[
\lambda_t=\push{(L_\alpha)}{\eta},
\qquad
\Bmap_t=K_{\alpha,\beta},
\qquad
\Smap_t=J_\gamma.
\]

For clarity, let $\mu^{\mathrm{pl}}_{t+h\mid t}(\boldsymbol{\theta})$ denote the plug-in forecast obtained by updating this common pre-update state with the fixed observation $\nu_t$ under candidate parameter vector $\boldsymbol{\theta}$, and subsequently propagating the updated state according to \eqref{eq:dhw-multistep-forecast}. This notation distinguishes the state-conditional forecast response considered below from the fitted forecasts used in estimation.

\begin{theorem}[Finite-horizon identification]
\label{thm:finite-horizon-identification}

Suppose $m\geq3$, the pre-update state is fixed, and $\eta$ has an atomless distribution with full support on $\R$. Assume that the relevant maps satisfy the map and moment conditions of Section~\ref{sec:dhw-transport-maps} and that the realised innovation is non-trivial:
\[
W_2(\nu_t,\mu_t)>0.
\]
Then, on the parameter domain
\[
0<\alpha\leq1,
\qquad
0\leq\beta,\gamma\leq1,
\]
the mapping
\[
\boldsymbol{\theta}
\longmapsto
\left\{
\mu^{\mathrm{pl}}_{t+1\mid t}(\boldsymbol{\theta}),
\mu^{\mathrm{pl}}_{t+2\mid t}(\boldsymbol{\theta}),
\mu^{\mathrm{pl}}_{t+m\mid t}(\boldsymbol{\theta})
\right\}
\]
is injective.

\end{theorem}

The mechanism behind Theorem~\ref{thm:finite-horizon-identification} follows directly from the timing of the seasonal updates. Let $F_h$ denote the increasing transport from $\eta$ to the corresponding state-conditional plug-in forecast. Since $m\geq3$, the first two forecast horizons use previously stored seasonal maps that are unaffected by the update at time $t$. Removing these maps gives
\[
H_1
=
\Smap_{t+1-m}^{-1}\circ F_1
=
K_{\alpha,\beta}\circ L_\alpha,
\]
and
\[
H_2
=
\Smap_{t+2-m}^{-1}\circ F_2
=
K_{\alpha,\beta}^{\circ2}\circ L_\alpha.
\]
Consequently,
\begin{equation}
K_{\alpha,\beta}
=
H_2\circ H_1^{-1},
\qquad
L_\alpha
=
K_{\alpha,\beta}^{-1}\circ H_1.
\label{eq:identify-level-trend}
\end{equation}
Thus the first two horizons separate the immediate level response from its subsequent propagation through the trend. Since $L_\alpha=\Id+\alpha \Delta_t^{\mathrm{pre}}$, a non-trivial innovation identifies $\alpha$. Moreover,
\[
K_{\alpha,\beta}\circ B^{-1}
=
\Id+\alpha\beta \Delta_t^{\mathrm{pre}}
\]
identifies $\alpha\beta$, and therefore $\beta$ because $\alpha>0$.

At horizon $m$, the seasonal position active at time $t$ is encountered again. The updated seasonal map $J_\gamma$ therefore enters the forecast for the first time, giving
\[
F_m
=
J_\gamma
\circ
K_{\alpha,\beta}^{\circ m}
\circ
L_\alpha.
\]
Once $\alpha$ and $\beta$ have been identified,
\begin{equation}
J_\gamma
=
F_m
\circ
\left(
K_{\alpha,\beta}^{\circ m}
\circ
L_\alpha
\right)^{-1}.
\label{eq:identify-seasonality}
\end{equation}
Since
\[
J_\gamma\circ S^{-1}
=
\Id+\gamma \Delta_t^{\mathrm{post}},
\]
the remaining parameter $\gamma$ is identified. The complete proof is given in Appendix~\ref{app:finite-horizon-identification}.

The theorem provides a structural identification result conditional on a common pre-update state and a realised innovation. It demonstrates how the level response, trend propagation, and seasonal recurrence distinguish the three smoothing parameters. It does not, by itself, establish global identification or consistency of the estimator in \eqref{eq:dhw-estimation-objective}, since candidate parameters generally generate different filtered states when applied recursively to an observed time series. Those statistical properties would require additional conditions on the resulting stochastic recursion.

The non-trivial innovation condition is also essential. If $\Emap=\Id$, then $\Delta_t^{\mathrm{post}}=\Delta_t^{\mathrm{pre}}=0$ and the current state update does not depend on the smoothing parameters. More generally, although the theorem establishes injectivity whenever the innovation is non-trivial, parameter estimation may be less informative when the realised innovation or its induced state response is small.


\section{Simulation Study}
\label{sec:simulation-study}

We use simulation to address two questions: whether adding trend and seasonality improves forecasts when those components are genuinely present in the underlying distributional dynamics, and how accurately the associated smoothing parameters can be recovered in finite samples. We therefore consider three nested data-generating processes (DGPs) that introduce level, trend, and seasonality sequentially. This design makes the source of any forecasting gain transparent.

\subsection{Simulation Design}
\label{subsec:simulation-design}

All data are generated directly from the DHW recursion, so that the true one-step-ahead conditional forecast distribution $\mu_t$ is available. We take the seasonal period to be $m=4$. In the \emph{Level} DGP we set $\alpha=0.35$ and $\beta=\gamma=0$; in the \emph{Level + trend} DGP we set $\alpha=0.35$, $\beta=0.30$, and $\gamma=0$; and in the \emph{Full DHW} DGP we set $\alpha=0.35$, $\beta=0.30$, and $\gamma=0.50$. Thus $\delta=\alpha\beta=0.105$ whenever the trend component is present. The reference distribution is a standardised non-Gaussian g-and-k distribution, while the non-identity trend and seasonal maps are nonlinear monotone transformations. Full DGP specifications are given in Appendix~\ref{app:simulation}.

The three DGPs form the structural hierarchy
\[
\text{Level}
\subset
\text{Level + trend}
\subset
\text{Full DHW}.
\]
In the Level DGP, both the trend and seasonal maps are identities. The Level + trend DGP additionally introduces a nonlinear trend map while retaining identity seasonal maps. The Full DHW DGP includes both nonlinear trend and seasonal components.

For each DGP, we consider two innovation mechanisms. The first uses translation innovations,
\[
\Emap_t(x)=x+\xi_t,
\]
where $\xi_t$ is centred. The second uses centred shape-changing innovations of the form
\[
\Emap_t(x)=x+\xi_t+\zeta_t g_t(x),
\]
where both $\xi_t$ and $\zeta_t$ are centred and $g_t$ is a smooth nonlinear function centred over the current forecast distribution. The second mechanism allows innovations to alter distributional shape rather than acting only through location. The innovation scales are chosen so that $\Emap_t$ remains strictly increasing almost surely.

We fit four models to each simulated series. WES contains only the distributional level component; WES-T additionally estimates a trend component; DHW with fixed seasonality (DHW-FS) estimates level and trend while retaining the seasonal maps obtained at initialisation as fixed; and DHW estimates all three smoothing parameters and updates the seasonal maps dynamically. Hence WES is correctly specified under the Level DGP, WES-T is correctly specified under the Level + trend DGP, and full DHW is correctly specified under the Full DHW DGP. DHW-FS is included to distinguish the value of a fixed seasonal adjustment from that of dynamically updated seasonality.

Each replication contains $500$ distributions, partitioned as
\[
40\text{ initialisation}
+60\text{ burn-in}
+300\text{ estimation}
+100\text{ test}.
\]
The first $40$ observations contain complete seasonal cycles and are used only to initialise the state. The next $60$ observations are filtered before evaluation of the estimation criterion, thereby reducing sensitivity to the initial state. Parameters are estimated from the subsequent $300$ observations, and forecasting performance is evaluated over the final $100$ observations. During the test period, parameters remain fixed but the filtered state is updated after each observed distribution, giving rolling one-step-ahead forecasts. We use $500$ Monte Carlo replications for each of the six DGP--innovation combinations.

The numerical representation uses $K=199$ probability-grid points and $M_x=4001$ map-grid points. Appendix~\ref{app:simulation} reports finite-sample and numerical sensitivity checks for these choices.

\subsection{Forecasting Performance}
\label{subsec:simulation-forecasting}

We evaluate one-step-ahead forecasts using two complementary criteria. The first is the out-of-sample mean squared Wasserstein error,
\[
\operatorname{MSWE}
=
\frac{1}{|\mathcal T_{\mathrm{test}}|}
\sum_{t\in\mathcal T_{\mathrm{test}}}
W_2^2\bigl(\nu_t,\widehat\mu_t\bigr),
\]
where $\widehat\mu_t$ is the estimated one-step-ahead forecast distribution. This is the practically relevant predictive loss, but it contains both irreducible innovation variation and error in estimating the conditional forecast.

Because the true DGP is known in simulation, we can isolate the latter component through the \emph{forecast-to-oracle error}
\[
\operatorname{FOE}
=
\frac{1}{|\mathcal T_{\mathrm{test}}|}
\sum_{t\in\mathcal T_{\mathrm{test}}}
W_2^2\bigl(\widehat\mu_t,\mu_t\bigr),
\]
where $\mu_t$ is the true conditional forecast distribution. FOE directly measures how far the fitted forecast is from the oracle conditional forecast and is the simulation analogue of the excess-risk term in Section~\ref{sec:theory}. We therefore use FOE as the primary measure for comparing structural specifications, while MSWE measures overall predictive performance. The oracle MSWE based on $\mu_t$ is also recorded as a diagnostic for the conditional risk decomposition.

\begin{figure}[t]
\centering
\includegraphics[width=0.98\textwidth]{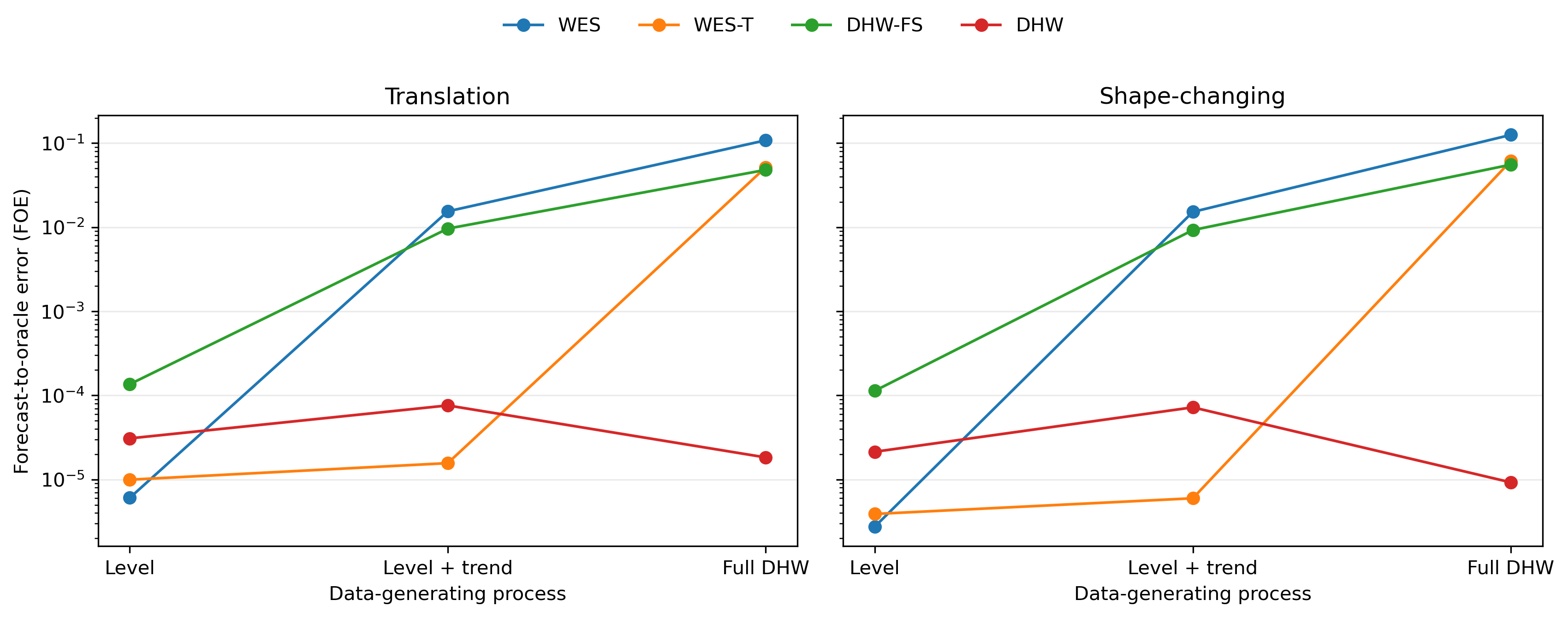}
\caption{One-step-ahead forecast-to-oracle error (FOE) across the nested simulation designs. Points show Monte Carlo means over $500$ replications. Lower values indicate forecasts closer to the true conditional forecast distribution. The vertical axis is logarithmic.}
\label{fig:simulation-foe}
\end{figure}

Figure~\ref{fig:simulation-foe} summarises FOE across the nested DGPs and fitted models, and Table~\ref{tab:simulation-forecast} reports the corresponding MSWE and FOE values. The same structural pattern appears under both innovation mechanisms. Under the Level DGP, WES gives the smallest FOE and WES-T is very close, so allowing an unnecessary trend carries little finite-sample cost. Full DHW also remains close to the oracle forecast in absolute terms, whereas DHW-FS performs less well because it retains a finite-sample seasonal pattern even though the true seasonal component is absent. Once trend is introduced, omitting it is costly: under the Level + trend DGP, WES has FOE around $1.5\times10^{-2}$ under both innovation mechanisms, compared with $1.6\times10^{-5}$ and $6.0\times10^{-6}$ for the correctly specified WES-T model. Full DHW remains accurate in absolute terms, although estimating an unnecessary adaptive seasonal component introduces a modest additional error.

The contrast is strongest under the Full DHW DGP. Full DHW has FOE $1.83\times10^{-5}$ under translation innovations and $9.24\times10^{-6}$ under shape-changing innovations. By comparison, WES-T and DHW-FS have FOEs between approximately $0.048$ and $0.061$, while WES exceeds $0.10$. Thus a fixed seasonal adjustment is not sufficient once the seasonal component itself evolves: dynamically updating the seasonal maps is essential for tracking the oracle conditional forecast. MSWE gives the same qualitative ordering.

The small absolute FOE values for the correctly specified models are a consequence of what this experiment is designed to isolate. Each observation is treated as an observed probability distribution, rather than as an estimate constructed from an additional finite sample, and the correctly specified DGP lies inside the fitted model class. The $K=199$ quantile ordinates are therefore a numerical representation of a single distribution, not $199$ independent observations. Across the six correctly specified DGP--innovation combinations, mean FOE is only about $0.2\%$--$1.1\%$ of the corresponding oracle MSWE. The grid-sensitivity experiment in Appendix~\ref{app:simulation} confirms that increasing the numerical resolution has negligible effect on these conclusions. Monte Carlo standard errors are reported in the appendix; they are small relative to the structural misspecification effects highlighted above.

\begin{table}[t]
\centering
\small
\caption{One-step-ahead forecasting performance. Entries are Monte Carlo means over $500$ replications and are multiplied by $10^3$. Lower values indicate better forecasting performance.}
\label{tab:simulation-forecast}
\begin{tabular}{llcccc}
\toprule
Innovation & DGP & WES & WES-T & DHW-FS & DHW \\
\midrule
\multicolumn{6}{l}{\textit{Panel A: $10^3\times$ MSWE}} \\
Translation & Level & 1.603 & 1.607 & 1.736 & 1.629 \\
 & Level + trend & 17.096 & 1.622 & 11.216 & 1.686 \\
 & Full DHW & 109.847 & 53.239 & 49.771 & 1.626 \\
Shape-changing & Level & 1.374 & 1.375 & 1.486 & 1.392 \\
 & Level + trend & 16.833 & 1.638 & 10.886 & 1.702 \\
 & Full DHW & 126.970 & 62.984 & 57.095 & 1.657 \\
\midrule
\multicolumn{6}{l}{\textit{Panel B: $10^3\times$ FOE}} \\
Translation & Level & 0.006 & 0.010 & 0.136 & 0.031 \\
 & Level + trend & 15.453 & 0.016 & 9.636 & 0.076 \\
 & Full DHW & 108.243 & 51.667 & 48.147 & 0.018 \\
Shape-changing & Level & 0.003 & 0.004 & 0.114 & 0.021 \\
 & Level + trend & 15.219 & 0.006 & 9.257 & 0.072 \\
 & Full DHW & 125.478 & 61.411 & 55.512 & 0.009 \\
\bottomrule
\end{tabular}
\end{table}

\subsection{Parameter Recovery}
\label{subsec:simulation-parameters}

We next examine parameter recovery under the correctly specified member of the nested model family. For the Level DGP we estimate $\alpha$ using WES; for the Level + trend DGP we estimate $\alpha$ and $\beta$ using WES-T; and for the Full DHW DGP we estimate $\alpha$, $\beta$, and $\gamma$ using DHW. Following the estimation procedure in Section~\ref{sec:estimation-forecasting}, optimisation is performed in terms of $(\alpha,\delta,\gamma)$, where
\[
\delta=\alpha\beta,
\]
and $\beta$ is recovered as $\delta/\alpha$.

Table~\ref{tab:simulation-parameters} reports the Monte Carlo mean and root mean squared error (RMSE) of each active parameter. The means are close to the true values under both innovation mechanisms, but the RMSEs show that finite-sample uncertainty is not negligible. The RMSE of $\widehat\alpha$ ranges from $0.032$ to $0.047$, while the RMSE of $\widehat\gamma$ in the Full DHW DGP is $0.049$ under translation innovations and $0.036$ under shape-changing innovations. Recovery of $\beta$ is less precise: its RMSE ranges from $0.046$ to $0.080$, so in the least favourable case it is about $27\%$ of the true value $0.30$. By contrast, the directly estimated product $\delta=\alpha\beta$ has RMSE between approximately $0.011$ and $0.020$. This supports the reparameterisation used for numerical estimation: the product governing the trend correction is more stably estimated than the ratio used to recover $\beta$.

The relatively precise recovery should again be interpreted in light of the simulation design: the distribution-valued observations themselves are observed without additional within-distribution sampling error. The probability grid controls numerical approximation of Wasserstein quantities; it does not multiply the effective sample size. Appendix~\ref{app:simulation} provides a separate sample-size experiment showing systematic improvement as the estimation sample increases, together with low-$\alpha$ and low-$\gamma$ configurations where recovery is more difficult.

\begin{table}[t]
\centering
\small
\caption{Finite-sample recovery of the active smoothing parameters. Entries are Monte Carlo mean (RMSE) over $500$ replications.}
\label{tab:simulation-parameters}
\begin{tabular}{llccc}
\toprule
DGP & Parameter & True value & Translation & Shape-changing \\
\midrule
Level & $\alpha$ & 0.350 & 0.344 (0.044) & 0.349 (0.034) \\
Level + trend & $\alpha$ & 0.350 & 0.346 (0.047) & 0.348 (0.033) \\
 & $\beta$ & 0.300 & 0.302 (0.074) & 0.303 (0.046) \\
 & $\delta$ & 0.105 & 0.102 (0.017) & 0.104 (0.011) \\
Full DHW & $\alpha$ & 0.350 & 0.348 (0.045) & 0.352 (0.032) \\
 & $\beta$ & 0.300 & 0.308 (0.080) & 0.300 (0.053) \\
 & $\gamma$ & 0.500 & 0.501 (0.049) & 0.501 (0.036) \\
 & $\delta$ & 0.105 & 0.105 (0.020) & 0.104 (0.014) \\
\bottomrule
\end{tabular}
\end{table}

\subsection{Multi-step Forecasting}
\label{subsec:simulation-multistep}

Finally, we check whether the simple plug-in multi-step forecast in Section~\ref{sec:estimation-forecasting} remains close to the corresponding conditional Fr\'echet mean. This is nontrivial beyond one step because future filtered states depend on future innovations, so iterating the current forecast recursion without simulating those innovations need not coincide exactly with averaging over all possible future state evolutions. Using the Full DHW DGP, true parameters and true filtered states, we simulate $100$ independent histories to a common forecast origin $t^\star=400$ and approximate the conditional Fr\'echet mean from $2000$ independent future continuations at horizons $h\in\{1,2,4\}$. The average squared Wasserstein discrepancy is about $0.8\times10^{-6}$ at $h=1$, where the population discrepancy is exactly zero and the observed value therefore measures nested Monte Carlo error, remains of the same order at $h=2$, and rises only to about $1.6\times10^{-6}$ at $h=4$. These discrepancies are roughly three orders of magnitude smaller than the one-step oracle MSWE. Thus, for the simulation setting and horizons considered here, iterating the plug-in forecast recursion provides a practically accurate approximation to the conditional Fr\'echet forecast. Full details and the corresponding figure are reported in Appendix~\ref{app:simulation}.


\section{Empirical Applications}
\label{sec:empirical-applications}

We apply distributional Holt--Winters (DHW) to hourly subway ridership in New York City, hourly electricity use in Portugal, and monthly retail turnover in Australia. At each date, the observation is a distribution across stations, clients, or state--industry series. We evaluate one-step forecasts of these distributions against four distributional models and two seasonal baselines.

\subsection{Data and evaluation}
\label{subsec:real-design}

At each date, the individual scalar observations receive equal weight in the empirical distribution. We represent its quantile function at $K=101$ equally spaced probability levels $u_k=(k-1/2)/K$, $k=1,\ldots,K$, using linear interpolation of the ordered observations. We use a weekly seasonal period of $m=168$ for the two hourly applications and a yearly period of $m=12$ for retail. For the hourly data, the seasonal maps describe each hour of the week, accommodating daily patterns and differences between days. Table~\ref{tab:real-design} gives the chronological split. The initial block is used for state initialisation and unscored filtering, the fitting block for parameter estimation, and the final block for forecast evaluation. Appendix~\ref{app:empirical} describes the small adjustments used to make tied quantiles strictly increasing, the calendar conventions and the precise initialisation windows.

\begin{table}[tbp]
\centering
\caption{Data and chronological evaluation periods.}
\label{tab:real-design}
\small
\setlength{\tabcolsep}{4pt}
\begin{tabular}{@{}lrrlll@{}}
\toprule
Application & Panel size & $m$ & Initial block & Fitting & Test \\
\midrule
NYC subway & 424 & 168 & 13 Jan.--Dec. 2017 & 2018 & 2019 \\
Electricity & 318 & 168 & 2 Jan.--Dec. 2012 & 2013 & 2014 \\
Retail & 133 & 12 & Apr. 1982--Dec. 1993 & 1994--2005 & 2006--2018 \\
\bottomrule
\end{tabular}
\par\smallskip
\begin{minipage}{\linewidth}\footnotesize
Each hourly test period contains 8,758 scored hours out of 8,760 calendar hours. Retail contains 156 scored months. Excluded clock-change hours retain their calendar positions and contribute no fitting or test loss.
\end{minipage}
\end{table}

All test forecasts are one step ahead. A forecast is formed before the target observation is used to update the model. Estimated coefficients remain fixed during testing. DHW updates its states as observations arrive, while the distributional benchmarks update their lagged inputs and retain their fitted means or maps. All methods use the same quantile grid and are scored on the same observed target dates. Figure~\ref{fig:real-paths} compares observed and successive one-step forecast quantiles over one complete seasonal cycle near the start of each test period.

Let $Q_t$ be the observed quantile function and $\widehat Q_{t\mid t-1}^{(a)}$ the forecast from method $a$. We use the discrete squared Wasserstein loss
\begin{equation}
 L_t^{(a)}=\frac{1}{K}\sum_{k=1}^{K}
 \left\{Q_t(u_k)-\widehat Q_{t\mid t-1}^{(a)}(u_k)\right\}^{2},
 \qquad
 \overline L_a=\frac{1}{|\mathcal T|}\sum_{t\in\mathcal T}L_t^{(a)},
 \label{eq:real-loss}
\end{equation}
where $\mathcal T$ contains the observed test dates. The percentage reduction in loss relative to benchmark $b$ is
\begin{equation}
 R_b=100\left(1-\frac{\overline L_{\mathrm{DHW}}}{\overline L_b}\right).
 \label{eq:real-reduction}
\end{equation}
A positive value favours DHW. This percentage refers to mean squared Wasserstein loss, not to the error at every date. Losses have different units across applications and should not be compared between datasets.

\subsection{DHW and benchmark specifications}
\label{subsec:real-models}

The empirical analysis uses the shared \texttt{dhwsmooth} implementation described in Appendix~\ref{app:dhwsmooth}. Initialisation follows Section~\ref{sec:dhw-initialisation}: the initial level is a Wasserstein barycentre, the trend map is the identity, and the seasonal maps transport the initial level to the average distribution at each seasonal position. Each candidate parameter vector is filtered through the remaining initial block before its fitting loss is computed. Parameters are estimated by constrained optimisation in $(\alpha,\delta,\gamma)$, with $\delta=\alpha\beta$. Value-map grids use pre-test dates; interpolation and unit-slope tail continuation follow Section~\ref{sec:dhw-numerical-implementation}. Appendix~\ref{app:empirical-numerics} reports the starting values, fitted coefficients and missing-hour updates.

We compare DHW with four nonseasonal distributional specifications: consecutive-transport ATM(1) \citep{atmzhu2023}, scalar WAR(1) \citep{zhang2022wasserstein}, the uniform-quantile variant DAM--UQ \citep{ghodrati2024distributional}, and lag-one DoDRM \citep{ghodrati2022distribution}. The cited formulations do not include an explicit seasonal component. Their estimation criteria and numerical conventions are specified in Appendix~\ref{app:empirical-benchmarks}. Two simple seasonal baselines use the latest available distribution at the same seasonal position and the expanding mean quantile function at that position. They are labelled seasonal naive and seasonal mean, respectively.

\begin{table}[tbp]
\centering
\caption{Mean one-step squared Wasserstein loss on the test periods.}
\label{tab:real-accuracy}
\small
\begin{tabular}{@{}lrrr@{}}
\toprule
Method & NYC & Electricity & Retail \\
\midrule
DHW & \textbf{6,491.07} & \textbf{7,937.34} & \textbf{0.00154} \\
Seasonal naive & 42,480.09 & 24,857.37 & 0.00433 \\
Seasonal mean & 25,566.36 & 44,964.29 & 0.62362 \\
ATM(1) & 123,937.14 & 22,085.51 & 0.01634 \\
WAR(1) & 129,526.69 & 26,667.03 & 0.02893 \\
DAM--UQ & 139,682.29 & 27,530.82 & 0.02220 \\
DoDRM & 120,425.02 & 26,958.66 & 0.01965 \\
\bottomrule
\end{tabular}
\par\smallskip
\begin{minipage}{\linewidth}\footnotesize
Smaller values are better. Bold values mark the smallest loss in each application. Retail losses use natural-log turnover. Benchmark labels refer to the specifications in Section~\ref{subsec:real-models} and Appendix~\ref{app:empirical-benchmarks}.
\end{minipage}
\end{table}

\subsection{New York City subway ridership}
\label{subsec:real-nyc}

The NYC panel contains 424 stations identified from the 2017--2018 records. Hourly ridership is summed across fare categories at each station. At an otherwise observed source hour, an absent station record is assigned zero. The target is the distribution of hourly ridership across these stations, including zero counts. Ridership is kept on its original scale. Figure~\ref{fig:real-paths}(a) shows the changing location and spread of this distribution over the first complete test week.

DHW has a mean test loss of 6,491.07 (Table~\ref{tab:real-accuracy}). Among the four distributional benchmarks, DoDRM has the lowest loss, at 120,425.02; DHW reduces this loss by 94.61\%. Relative to seasonal naive and seasonal mean, the reductions are 84.72\% and 74.61\%, respectively. The fitted trend-smoothing coefficient is close to zero, $\widehat\beta=1.35\times10^{-11}$, so the trend map changes very little as new observations arrive.

The mean-loss gain is not uniform across hours. DHW has lower mean loss than seasonal naive in 145 of the 168 hour-of-week groups, but lower date-level loss in 4,371 of the 8,758 scored hours. Its median date-level loss is 1,507.75, compared with 1,705.80 for seasonal naive. Although DHW has a much smaller average loss, its relative performance varies considerably across individual hours.

\subsection{Portuguese electricity use}
\label{subsec:real-uci}

The electricity data come from \textit{ElectricityLoadDiagrams20112014} \citep{electricityloaddiagrams20112014_321}. We retain 318 clients with positive use by the end of 1 January 2012 and at least one positive reading in December 2013. These selection dates precede the test period. Four quarter-hour power readings are summed and divided by four to obtain hourly energy use in kWh. Ordinary zero readings are retained. Each hourly distribution describes differences in use across the retained clients.

DHW has a mean test loss of 7,937.34. ATM(1), the best of the four distributional benchmarks on this test period, has a loss of 22,085.51, giving a reduction of 64.06\%. DHW also reduces the seasonal-naive loss by 68.07\% and has lower mean loss in 150 of the 168 hour-of-week groups. Figure~\ref{fig:real-paths}(b) shows forecasts tracking the daily variation in both the median and upper quantiles. The fitted trend coefficient is small but positive, $\widehat\beta=0.0005$, while the seasonal coefficient is $\widehat\gamma=0.0325$.

\subsection{Australian retail turnover}
\label{subsec:real-retail}

The retail data are Australian Bureau of Statistics series in the \texttt{tsibbledata} package\footnote{\url{https://tsibbledata.tidyverts.org/reference/aus_retail.html}}. We retain 133 state--industry series with complete, positive turnover from April 1982 to December 2018. These published series include both broad industry aggregates and their constituent categories, with each series receiving equal weight. Turnover is measured in AUD millions. We take the natural logarithm before computing cross-sectional quantiles, so the target is the distribution of log turnover across these series. The complete-panel rule uses test-period information; the reported results describe this retrospectively selected panel.

\begin{figure}[p]
\centering
\includegraphics[width=\linewidth,height=0.76\textheight,keepaspectratio]{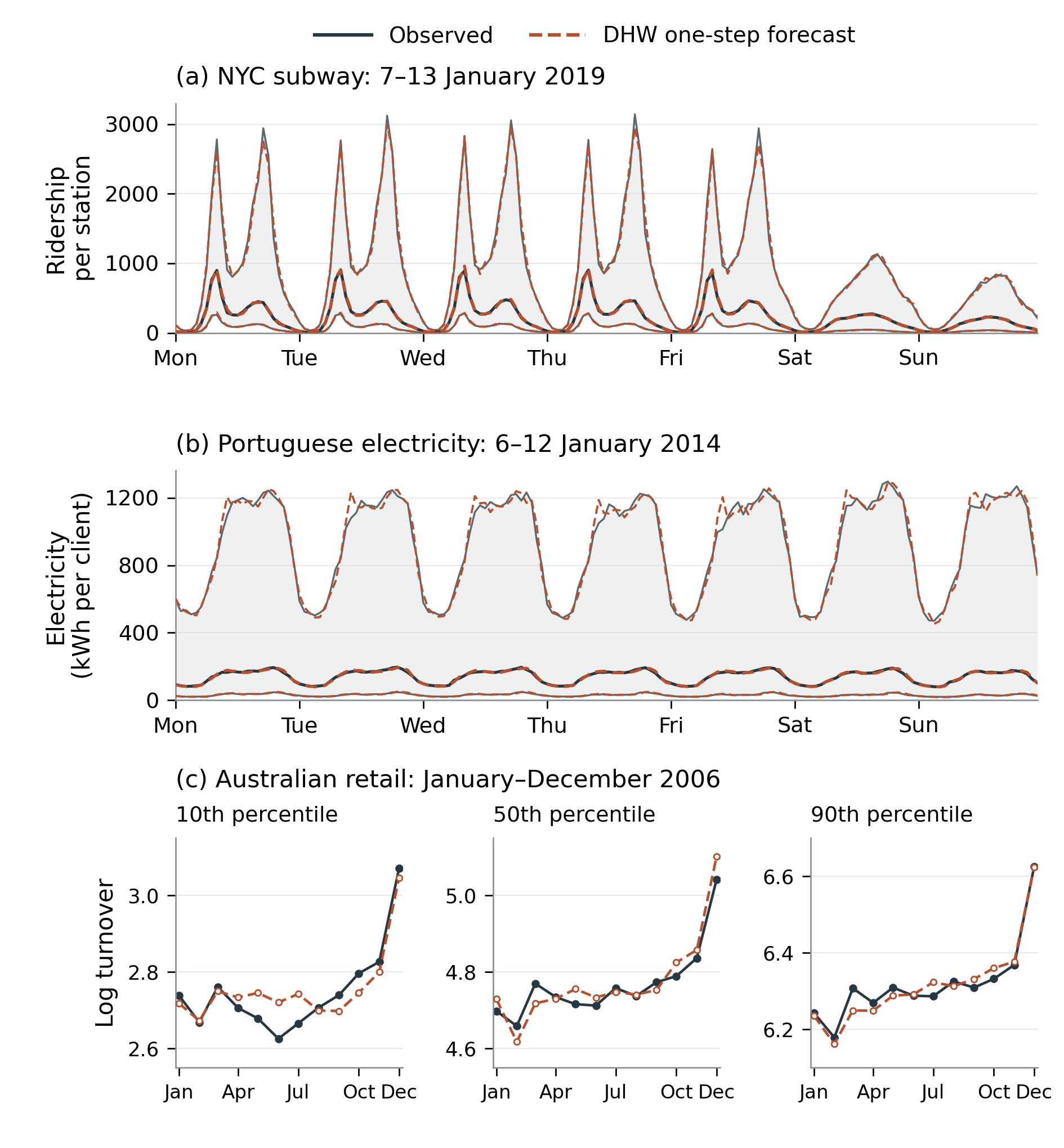}
\caption{Observed and rolling one-step DHW forecast quantiles over one complete seasonal cycle: 7--13 January 2019 for NYC, 6--12 January 2014 for electricity, and January--December 2006 for retail. Each forecast uses the observations available before its target date. The hourly windows are the first complete Monday--Sunday test weeks; the lines show the 10th, 50th and 90th percentiles, and shading spans the observed 10th--90th percentiles. The retail quantiles are shown separately with equal vertical spans of 0.6 log units and different vertical origins. All quantiles describe cross-sectional variation across stations, clients or retail series.}
\label{fig:real-paths}
\end{figure}

DHW has a mean test loss of 0.00154 in squared log units, compared with 0.01634 for ATM(1), the lowest-loss distributional benchmark. The reduction is 90.59\%. DHW also improves on seasonal naive by 64.52\%, with lower mean loss in every calendar-month group and every test year. The seasonal mean performs poorly over the long test period: averaging all past observations at a given month responds slowly to changes in the distribution's level. Figure~\ref{fig:real-paths}(c) shows the February decline and December peak in the first test year. DHW follows these movements at all three displayed quantiles, with larger deviations at the 10th percentile in the middle of the year.

\begin{figure}[p]
\centering
\includegraphics[width=\linewidth,height=0.75\textheight,keepaspectratio]{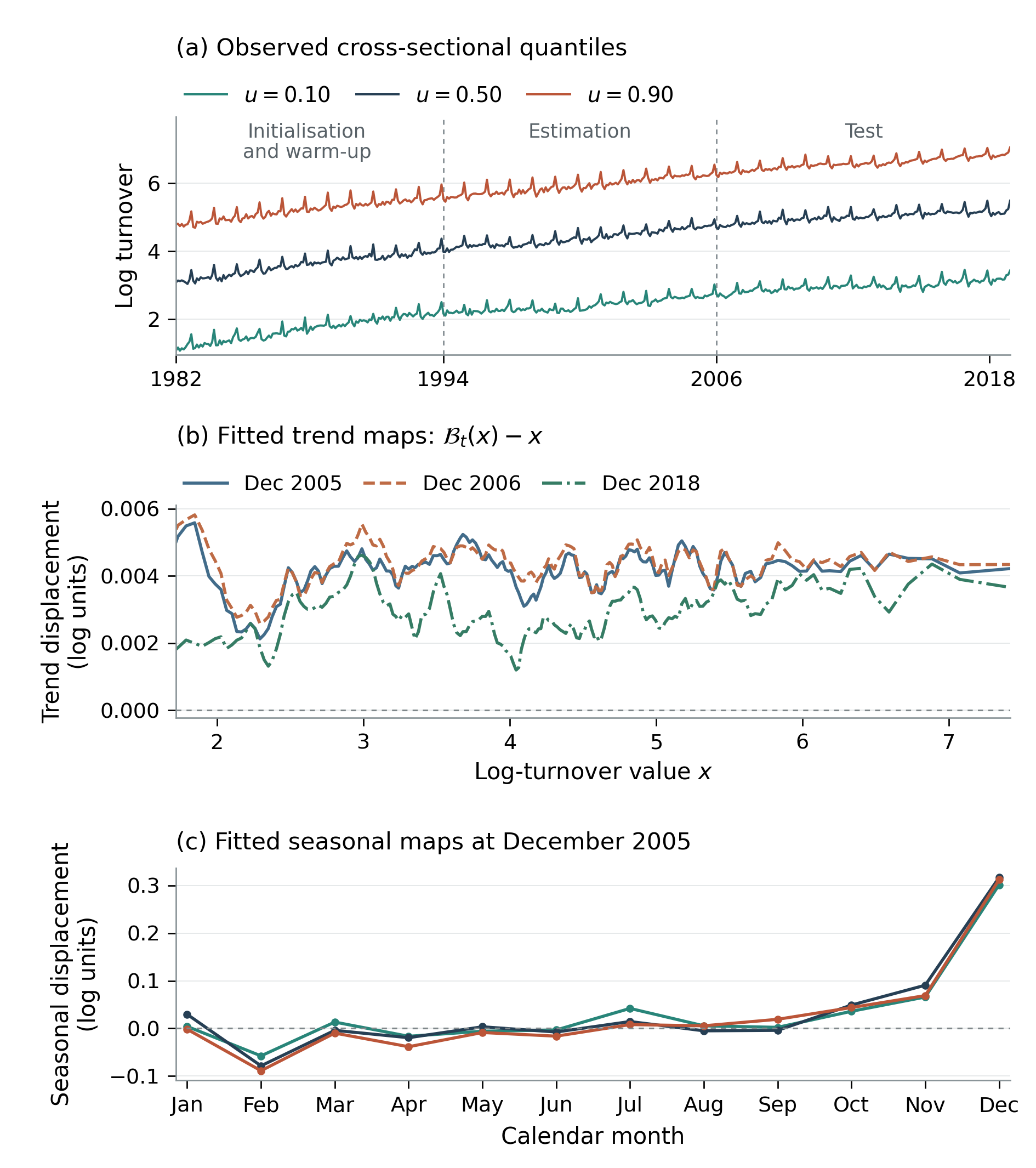}
\caption{Observed distributions and fitted DHW components for Australian retail turnover. (a) Observed 10th, 50th and 90th percentiles, with the initial, fitting and test blocks indicated. (b) Trend displacements $\Bmap_t(x)-x$ after December 2005, December 2006 and December 2018, evaluated on a common value range. (c) The twelve seasonal maps stored after December 2005, evaluated at the 10th, 50th and 90th percentiles of the January 2006 trend-adjusted distribution. The reference values are held fixed across months. Displacements are measured in natural-log turnover units; the zero line denotes an identity effect. Appendix~\ref{app:empirical-components} defines the evaluations precisely.}
\label{fig:retail-anatomy}
\end{figure}

Figure~\ref{fig:retail-anatomy} relates the fitted maps to the observed distributional trajectory. Panel (a) shows the observed quantiles through time, while panel (b) displays $\Bmap_t(x)-x$ at three filtered states. These trend displacements are positive over the displayed range and vary with $x$. Panel (c) evaluates the complete seasonal cycle stored after December 2005 at three fixed quantiles of the next trend-adjusted distribution. Differences between the curves show how seasonal effects vary across the distribution. Their crossings indicate changes in the relative size of these effects across quantiles. Each seasonal map remains increasing and preserves the ordering of the transformed quantiles. The corresponding seasonal maps after December 2006 and December 2018 are compared on the same reference points in Appendix~\ref{app:empirical-components}.

\subsection{Paired comparisons}
\label{subsec:real-inference}

Table~\ref{tab:real-inference} reports percentage loss reductions and approximate pointwise 95\% intervals from 19,999 paired circular block-bootstrap resamples \citep{politis1992circular}. Blocks contain 168 calendar hours for NYC and electricity and 12 months for retail. The same sampled dates are used for both methods, with unavailable calendar positions retained. Two-sided tests of equal expected loss use centred bootstrap loss differences; Holm's procedure adjusts the $p$-values across the twelve comparisons. Appendix~\ref{app:empirical-uncertainty} gives further details.

\begin{table}[tbp]
\centering
\caption{DHW loss reduction (\%) relative to each distributional benchmark, with approximate pointwise 95\% block-bootstrap intervals.}
\label{tab:real-inference}
\small
\setlength{\tabcolsep}{5pt}
\begin{tabular}{@{}lccc@{}}
\toprule
Benchmark & NYC & Electricity & Retail \\
\midrule
ATM(1) & 94.76 [93.63, 95.65] & 64.06 [61.79, 66.38] & 90.59 [89.95, 91.20] \\
WAR(1) & 94.99 [93.91, 95.84] & 70.24 [68.64, 71.93] & 94.68 [94.02, 95.27] \\
DAM--UQ & 95.35 [94.35, 96.14] & 71.17 [69.59, 72.84] & 93.08 [92.44, 93.66] \\
DoDRM & 94.61 [93.46, 95.52] & 70.56 [69.09, 72.12] & 92.18 [91.42, 92.87] \\
\bottomrule
\end{tabular}
\end{table}

All twelve comparisons favour DHW, with Holm-adjusted bootstrap $p$-values below 0.001. Inference is conditional on the fitted models and the stations, clients and retail series included in the analysis. DHW also has lower mean loss than each distributional benchmark in every calendar month of the hourly test periods and every year of the retail test period. The 95\% intervals remain above zero when the blocks span two or four seasonal cycles (Appendix~\ref{app:empirical-uncertainty}). These results provide consistent evidence of improved one-step forecast accuracy across the three applications.

\FloatBarrier

\section{Conclusion}
\label{sec:conclusion}

We have developed a distributional Holt--Winters method that represents level by a probability distribution and trend and seasonality by monotone transport maps. The construction retains the recursive error-correction structure of classical exponential smoothing while allowing changes in location, dispersion and tail behaviour. Under the stated map and moment conditions, the updates preserve valid distributions without a separate projection. The quantile representation provides a direct numerical implementation, and estimation uses one-step squared Wasserstein prediction loss. Conditional centring of the innovations gives this criterion a statistical interpretation through the one-step risk decomposition and conditional Fr\'echet optimality result.

The simulation study separates the contributions of the components. Simpler specifications perform well when their restrictions match the data-generating process, while updating the seasonal maps is important when seasonal effects evolve. The empirical applications provide complementary evidence: DHW has lower mean one-step loss than the distributional benchmarks and simple seasonal baselines considered for subway ridership, electricity use and retail turnover. The retail component plots show how fitted trend and seasonal effects vary across the distribution. The component maps offer an interpretable description alongside the forecast-accuracy results.

The present framework focuses on univariate settings. Its one-step optimality result does not extend automatically to nonlinear multi-step plug-in forecasts, although the discrepancy from conditional Fr\'echet forecasts is small in the simulation settings studied. Similarly, the finite-horizon identification result holds conditional on a common pre-update state; it does not establish global identification or consistency of the fitted parameters. Statistical analysis of the full filtered recursion and methods for quantifying uncertainty about future distributions are natural directions for further work.


\bibliographystyle{abbrvnat}
\bibliography{ref_dhw}

@article{matsubara2026WES,
  title={Wasserstein exponential smoothing},
  author={Takuo Matsubara and Peiwen Jiang and Minh-Ngoc Tran and Wilson Ye Chen},
  journal={Working paper},
  year={2026},
}

@article{panaretos2019statistical,
  title={Statistical Aspects of {Wasserstein} Distances},
  author={Panaretos, Victor M. and Zemel, Yoav},
  journal={Annual Review of Statistics and Its Application},
  volume={6},
  pages={405--431},
  year={2019},
  doi={10.1146/annurev-statistics-030718-104938}
}

@article{kokoszka2019forecasting,
  title={Forecasting of Density Functions with an Application to Cross-Sectional and Intraday Returns},
  author={Kokoszka, Piotr and Miao, Hong and Petersen, Alexander and Shang, Han Lin},
  journal={International Journal of Forecasting},
  volume={35},
  number={4},
  pages={1304--1317},
  year={2019},
  doi={10.1016/j.ijforecast.2019.05.007}
}

@article{zhang2022wasserstein,
  title={{Wasserstein} Autoregressive Models for Density Time Series},
  author={Zhang, Chao and Kokoszka, Piotr and Petersen, Alexander},
  journal={Journal of Time Series Analysis},
  volume={43},
  number={1},
  pages={30--52},
  year={2022},
  doi={10.1111/jtsa.12590}
}

@article{holt2004forecasting,
  title={Forecasting Seasonals and Trends by Exponentially Weighted Moving Averages},
  author={Holt, Charles C.},
  journal={International Journal of Forecasting},
  volume={20},
  number={1},
  pages={5--10},
  year={2004},
  doi={10.1016/j.ijforecast.2003.09.015}
}

@article{winters1960forecasting,
  title={Forecasting Sales by Exponentially Weighted Moving Averages},
  author={Winters, Peter R.},
  journal={Management Science},
  volume={6},
  number={3},
  pages={324--342},
  year={1960},
  doi={10.1287/mnsc.6.3.324}
}

@article{hyndman2002state,
  title={A State Space Framework for Automatic Forecasting Using Exponential Smoothing Methods},
  author={Hyndman, Rob J. and Koehler, Anne B. and Snyder, Ralph D. and Grose, Simone},
  journal={International Journal of Forecasting},
  volume={18},
  number={3},
  pages={439--454},
  year={2002},
  doi={10.1016/S0169-2070(01)00110-8}
}

@book{hyndman2008forecasting,
  title={Forecasting with Exponential Smoothing: The State Space Approach},
  author={Hyndman, Rob J. and Koehler, Anne B. and Ord, J. Keith and Snyder, Ralph D.},
  series={Springer Series in Statistics},
  publisher={Springer},
  address={Berlin},
  year={2008},
  doi={10.1007/978-3-540-71918-2}
}

@misc{electricityloaddiagrams20112014_321,
  author       = {Trindade, Artur},
  title        = {{ElectricityLoadDiagrams20112014}},
  year         = {2015},
  howpublished = {UCI Machine Learning Repository},
  note         = {{DOI}: https://doi.org/10.24432/C58C86}
}

@article{atmzhu2023,
author = {Zhu, Changbo and Müller, Hans-Georg},
year = {2023},
month = {05},
pages = {1012-1033},
title = {Autoregressive optimal transport models},
volume = {85},
journal = {Journal of the Royal Statistical Society Series B: Statistical Methodology},
doi = {10.1093/jrsssb/qkad051}
}

@article{ghodrati2022distribution,
  title={Distribution-on-distribution regression via optimal transport maps},
  author={Ghodrati, Laya and Panaretos, Victor M},
  journal={Biometrika},
  volume={109},
  number={4},
  pages={957--974},
  year={2022},
  publisher={Oxford University Press}
}

@article{ghodrati2024distributional,
  title={On distributional autoregression and iterated transportation},
  author={Ghodrati, Laya and Panaretos, Victor M},
  journal={Journal of Time Series Analysis},
  volume={45},
  number={5},
  pages={739--770},
  year={2024},
  publisher={Wiley Online Library}
}

@incollection{politis1992circular,
  author={Politis, Dimitris N. and Romano, Joseph P.},
  title={A Circular Block-Resampling Procedure for Stationary Data},
  booktitle={Exploring the Limits of Bootstrap},
  editor={LePage, Raoul and Billard, Lynne},
  pages={263--270},
  publisher={John Wiley},
  year={1992}
}


\appendix
\crefalias{section}{appendix}
\crefalias{subsection}{appendix}
\crefalias{subsubsection}{appendix}


\section{Finite-Horizon Identification}
\label{app:finite-horizon-identification}

\begin{proof}[Proof of Theorem~\ref{thm:finite-horizon-identification}]

Fix the pre-update state at time $t$ and the realised observation $\nu_t$. Throughout the proof, write
\[
B=\Bmap_{t-1},
\qquad
S=\Smap_{t-m},
\qquad
\eta=\eta_t,
\qquad
\Emap=\Emap_t.
\]
Recall that the pre-seasonal and post-seasonal innovation displacements are
\[
\Delta_t^{\mathrm{pre}}=S^{-1}\circ \Emap\circ S-\Id,
\qquad
\Delta_t^{\mathrm{post}}=\Emap-\Id.
\]
Since the pre-update state and $\nu_t$ are fixed, these maps are common to all candidate parameter vectors under consideration.

For $\boldsymbol{\theta}=(\alpha,\beta,\gamma)$, define
\[
\begin{aligned}
L_\alpha&=\Id+\alpha\Delta_t^{\mathrm{pre}},\\
K_{\alpha,\beta}&=(\Id+\alpha\beta\Delta_t^{\mathrm{pre}})\circ B,\\
J_\gamma&=(\Id+\gamma\Delta_t^{\mathrm{post}})\circ S.
\end{aligned}
\]
We first verify that these maps give the updated DHW state.

By the level update \eqref{eq:map-level-update} and the transport identity \eqref{eq:pullback-innovation},
\[
\lambda_t
=
\push{(L_\alpha)}{\eta}.
\]
Since $\eta=\push{(B)}{\lambda_{t-1}}$ and both $B$ and $L_\alpha$ are increasing, the transport from $\lambda_{t-1}$ to $\lambda_t$ is
\[
\otmap{\lambda_{t-1}}{\lambda_t}
=
L_\alpha\circ B
\]
up to the usual almost-everywhere identification of transport maps. Substituting into the trend update gives
\[
\begin{aligned}
\Bmap_t
&=
(1-\beta)B+\beta(L_\alpha\circ B)\\
&=
\left\{
(1-\beta)\Id+\beta L_\alpha
\right\}\circ B\\
&=
(\Id+\alpha\beta \Delta_t^{\mathrm{pre}})\circ B\\
&=
K_{\alpha,\beta}.
\end{aligned}
\]
Similarly, the optimal transport from $\eta$ to the observation is $\Emap\circ S$, so the seasonal update satisfies
\[
\begin{aligned}
\Smap_t
&=
(1-\gamma)S+\gamma(\Emap\circ S)\\
&=
\left\{
(1-\gamma)\Id+\gamma \Emap
\right\}\circ S\\
&=
(\Id+\gamma \Delta_t^{\mathrm{post}})\circ S\\
&=
J_\gamma.
\end{aligned}
\]
Under the stated full-support and continuity assumptions, these transport identities extend from almost-everywhere identities to identities of the corresponding continuous maps.

We next consider the three forecast horizons in the theorem. Since $\eta$ is atomless and has full support, and all maps in the forecast construction are strictly increasing continuous bijections, each forecast distribution uniquely determines its increasing transport from $\eta$. In particular, equality of two forecast distributions implies equality of their corresponding increasing transport maps.

For a candidate parameter vector $\boldsymbol{\theta}$, let $F_h(\boldsymbol{\theta})$ denote the increasing transport from $\eta$ to the state-conditional plug-in forecast $\mu^{\mathrm{pl}}_{t+h\mid t}(\boldsymbol{\theta})$.

From the seasonal-index convention in \eqref{eq:dhw-seasonal-forecast-index}, the first two forecast horizons use the seasonal maps
\[
G_1=\Smap_{t+1-m},
\qquad
G_2=\Smap_{t+2-m}.
\]
Since $m\geq3$, both maps belong to the fixed pre-update state and are unaffected by the observation at time $t$. Therefore,
\[
F_1(\boldsymbol{\theta})
=
G_1\circ K_{\alpha,\beta}\circ L_\alpha
\]
and
\[
F_2(\boldsymbol{\theta})
=
G_2\circ K_{\alpha,\beta}^{\circ2}\circ L_\alpha.
\]

Suppose two parameter vectors
\[
\boldsymbol{\theta}
=
(\alpha,\beta,\gamma)
\]
and
\[
\boldsymbol{\theta}'
=
(\alpha',\beta',\gamma')
\]
produce identical plug-in forecasts at horizons $1$, $2$, and $m$. Equality of the first two forecast distributions implies
\[
F_1(\boldsymbol{\theta})
=
F_1(\boldsymbol{\theta}')
\]
and
\[
F_2(\boldsymbol{\theta})
=
F_2(\boldsymbol{\theta}').
\]

Remove the known seasonal maps by defining
\[
H_1=G_1^{-1}\circ F_1,
\qquad
H_2=G_2^{-1}\circ F_2.
\]
It follows that
\[
H_1=K_{\alpha,\beta}\circ L_\alpha
\]
and
\[
H_2=K_{\alpha,\beta}^{\circ2}\circ L_\alpha.
\]
Because all maps involved are bijections,
\[
\begin{aligned}
H_2\circ H_1^{-1}
&=
K_{\alpha,\beta}^{\circ2}
\circ L_\alpha
\circ L_\alpha^{-1}
\circ K_{\alpha,\beta}^{-1}\\
&=
K_{\alpha,\beta}.
\end{aligned}
\]
Consequently, the first two forecast distributions uniquely determine
\[
K_{\alpha,\beta}
=
H_2\circ H_1^{-1}
\]
and subsequently
\[
L_\alpha
=
K_{\alpha,\beta}^{-1}\circ H_1.
\]
In particular,
\[
K_{\alpha,\beta}
=
K_{\alpha',\beta'}
\]
and
\[
L_\alpha=L_{\alpha'}.
\]

We now identify $\alpha$ and $\beta$. By definition,
\[
L_\alpha-L_{\alpha'}
=
(\alpha-\alpha')\Delta_t^{\mathrm{pre}}.
\]
The non-triviality condition
\[
W_2(\nu_t,\mu_t)>0
\]
implies that $\Emap$ differs from $\Id$ on a set of positive $\mu_t$-measure. Since
\[
\mu_t=\push{(S)}{\eta}
\]
and $S$ is a bijection, the conjugated innovation $S^{-1}\circ \Emap\circ S$ differs from $\Id$ on a set of positive $\eta$-measure. Hence $\Delta_t^{\mathrm{pre}}$ is not identically zero.

The equality $L_\alpha=L_{\alpha'}$ therefore implies
\[
\alpha=\alpha'.
\]

Next,
\[
K_{\alpha,\beta}
=
(\Id+\alpha\beta \Delta_t^{\mathrm{pre}})\circ B.
\]
Since $B$ is invertible,
\[
K_{\alpha,\beta}\circ B^{-1}
=
\Id+\alpha\beta \Delta_t^{\mathrm{pre}}.
\]
Similarly,
\[
K_{\alpha',\beta'}\circ B^{-1}
=
\Id+\alpha'\beta'\Delta_t^{\mathrm{pre}}.
\]
Equality of the two trend maps consequently gives
\[
(\alpha\beta-\alpha'\beta')\Delta_t^{\mathrm{pre}}=0.
\]
Because $\Delta_t^{\mathrm{pre}}$ is not identically zero,
\[
\alpha\beta=\alpha'\beta'.
\]
Together with $\alpha=\alpha'>0$, this implies
\[
\beta=\beta'.
\]

It remains to identify $\gamma$. At horizon $m$, the seasonal position corresponding to time $t$ is encountered again. According to \eqref{eq:dhw-seasonal-forecast-index}, its most recently updated seasonal map is $\Smap_t=J_\gamma$. Hence the increasing transport from $\eta$ to the $m$-step plug-in forecast is
\[
F_m(\boldsymbol{\theta})
=
J_\gamma
\circ
K_{\alpha,\beta}^{\circ m}
\circ
L_\alpha.
\]

The preceding argument has established
\[
\alpha=\alpha',
\qquad
\beta=\beta',
\]
and therefore
\[
L_\alpha=L_{\alpha'},
\qquad
K_{\alpha,\beta}=K_{\alpha',\beta'}.
\]
Since the two parameter vectors produce the same $m$-step forecast distribution, uniqueness of the increasing transport from $\eta$ gives
\[
F_m(\boldsymbol{\theta})
=
F_m(\boldsymbol{\theta}').
\]
Consequently,
\[
\begin{aligned}
J_\gamma
&=
F_m
\circ
\left(
K_{\alpha,\beta}^{\circ m}
\circ
L_\alpha
\right)^{-1}\\
&=
J_{\gamma'}.
\end{aligned}
\]

By definition,
\[
J_\gamma
=
(\Id+\gamma \Delta_t^{\mathrm{post}})\circ S.
\]
Thus
\[
J_\gamma\circ S^{-1}
=
\Id+\gamma \Delta_t^{\mathrm{post}}
\]
and
\[
J_{\gamma'}\circ S^{-1}
=
\Id+\gamma'\Delta_t^{\mathrm{post}}.
\]
Equality of the updated seasonal maps therefore implies
\[
(\gamma-\gamma')\Delta_t^{\mathrm{post}}=0.
\]

Again, the non-triviality condition $W_2(\nu_t,\mu_t)>0$ implies that $\Emap$ differs from $\Id$ on a set of positive $\mu_t$-measure. Hence $\Delta_t^{\mathrm{post}}=\Emap-\Id$ is not identically zero, and we obtain
\[
\gamma=\gamma'.
\]

We have therefore established that equality of the state-conditional plug-in forecasts at horizons $1$, $2$, and $m$ implies
\[
\alpha=\alpha',
\qquad
\beta=\beta',
\qquad
\gamma=\gamma'.
\]
Thus
\[
\boldsymbol{\theta}=\boldsymbol{\theta}',
\]
proving the asserted injectivity.

\end{proof}


\section{Additional Simulation Details and Robustness Checks}
\label{app:simulation}

\paragraph{DGP specification and numerical implementation.}
The reference quantile function is obtained from a g-and-k distribution. Let $z_u=\Phi^{-1}(u)$ and define
\[
\widetilde Q_0(u)
=
\left\{
1+0.8\tanh\left(\frac{0.4z_u}{2}\right)
\right\}
(1+z_u^2)^{0.2}z_u.
\]
We standardise $\widetilde Q_0$ numerically on the probability grid to obtain a reference quantile function $Q_0$ with zero mean and unit variance. The probability grid consists of $K=199$ equally spaced points on $[0.001,0.999]$.

When the trend component is active, its initial map is
\[
\Bmap_0(x)
=
x+0.03+0.05\tanh(x/1.2).
\]
When seasonality is active, the four initial seasonal maps are
\[
\Smap_{-j}(x)
=
x+a_j+r_j\tanh(x),
\qquad
j=0,1,2,3,
\]
where
\[
(a_0,a_1,a_2,a_3)
=
(-0.18,0.06,0.15,-0.03)
\]
and
\[
(r_0,r_1,r_2,r_3)
=
(0.08,-0.03,-0.05,0).
\]
Since $\sum_{j=0}^{3}a_j=\sum_{j=0}^{3}r_j=0$, the initial seasonal maps satisfy the required normalisation. In a DGP where a structural component is inactive, the corresponding map is set to $\Id$.

For translation innovations we use
\[
\Emap_t(x)
=
x+\xi_t,
\qquad
\xi_t\sim N(0,0.04^2).
\]
For shape-changing innovations we use
\[
\Emap_t(x)
=
x+\xi_t+\zeta_t g_t(x),
\qquad
\xi_t\sim N(0,0.03^2),
\qquad
\zeta_t\sim\operatorname{Unif}(-0.06,0.06),
\]
independently, where
\[
g_t(x)
=
\tanh\left(
\frac{x-c_t}{0.8}
\right)
-
\int_0^1
\tanh\left(
\frac{\qf{\mu_t}(u)-c_t}{0.8}
\right)\,du
\]
and $c_t$ is the median of $\mu_t$. The centring separates the nonlinear shape perturbation from a pure location shift. Moreover,
\[
\inf_x \Emap_t'(x)
\geq
1-\frac{0.06}{0.8}
>0,
\]
so the innovation map remains strictly increasing.

Transport maps are represented on $M_x=4001$ equally spaced points on $[-100,100]$. Between grid points, displacement is linearly interpolated; outside the grid, the boundary displacement is continued with unit slope. Smoothing parameters are estimated by minimising the one-step-ahead squared Wasserstein loss. The constrained optimisation is carried out in $(\alpha,\delta,\gamma)$ with
\[
0<\alpha\leq1,\qquad
0\leq\delta\leq\alpha,\qquad
0\leq\gamma\leq1.
\]
For restricted benchmark models, inactive parameters are fixed at zero and the corresponding maps are identities, except for DHW-FS, for which the seasonal maps are estimated at initialisation and subsequently held fixed.

\paragraph{Finite-sample parameter recovery.}
To examine how parameter recovery changes with the estimation sample size, we repeat the correctly specified Full DHW experiment under shape-changing innovations for
\[
N\in\{100,300,600,1200\},
\]
while retaining the same initialisation and burn-in lengths. In addition to the baseline parameter setting
\[
(\alpha,\beta,\gamma)
=
(0.35,0.30,0.50),
\]
we consider a low-$\alpha$ setting
\[
(\alpha,\beta,\gamma)
=
(0.10,0.30,0.50)
\]
and a low-$\gamma$ setting
\[
(\alpha,\beta,\gamma)
=
(0.35,0.30,0.10).
\]
The same simulated path prefixes are used across sample sizes within each Monte Carlo replication, and each parameter setting uses $100$ replications.

\begin{figure}[t]
\centering
\includegraphics[width=0.98\textwidth]{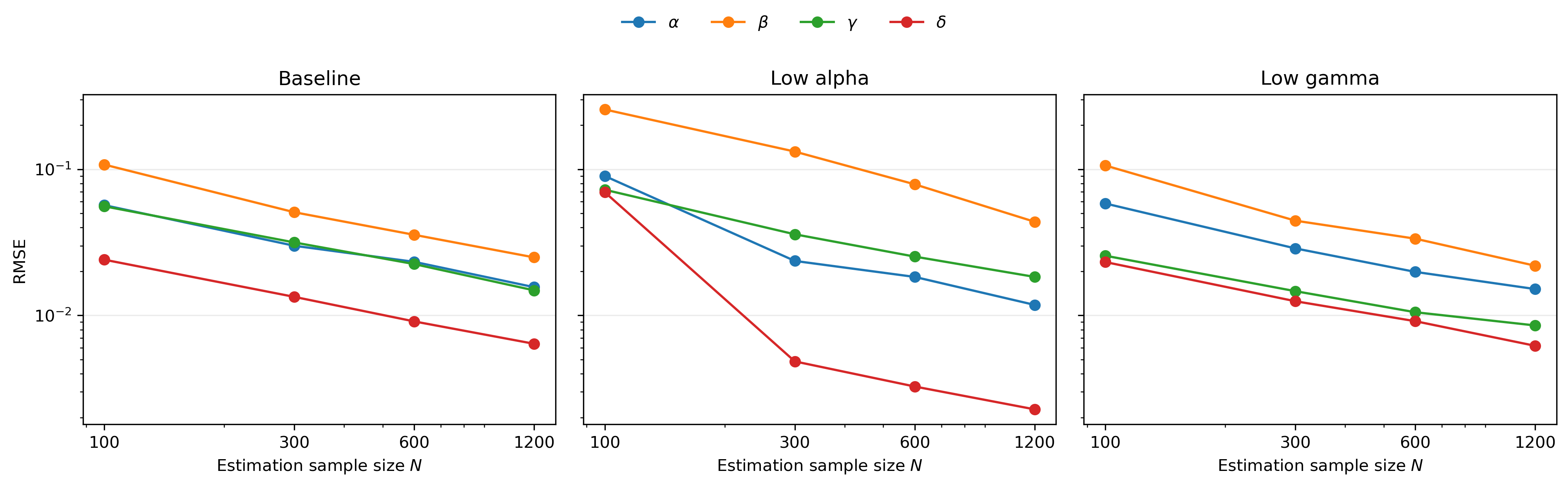}
\caption{Finite-sample parameter recovery as the estimation sample size increases. The figure reports RMSE over $100$ Monte Carlo replications under the baseline, low-$\alpha$, and low-$\gamma$ parameter settings.}
\label{fig:appendix-parameter-convergence}
\end{figure}

Figure~\ref{fig:appendix-parameter-convergence} shows a systematic decline in RMSE as the estimation sample grows. Under the baseline setting, the RMSEs of $\widehat\alpha$, $\widehat\beta$, $\widehat\gamma$, and $\widehat\delta$ fall from approximately $0.057$, $0.108$, $0.056$, and $0.024$ at $N=100$ to $0.016$, $0.025$, $0.015$, and $0.006$ at $N=1200$. The low-$\gamma$ setting displays a similarly regular improvement. The low-$\alpha$ setting is more difficult, especially for recovering $\beta$: its RMSE falls from $0.256$ at $N=100$ to $0.044$ at $N=1200$. In the same setting, $\delta$ is substantially more stable once the sample reaches the estimation size used in Section~\ref{sec:simulation-study}, with RMSE $0.0048$ at $N=300$ and $0.0023$ at $N=1200$. This reinforces the numerical motivation for estimating the product $\delta=\alpha\beta$ directly.

\paragraph{Multi-step plug-in approximation.}
The one-step forecast is the conditional Fr\'echet mean under the innovation-centering condition in Section~\ref{sec:theory}. Beyond one step, however, future filtered states depend on future innovations. Consequently, the convenient plug-in forecast obtained by iterating the current trend and seasonal maps need not coincide exactly with the conditional Fr\'echet mean obtained after averaging over all future innovation paths.

We examine the size of this discrepancy using the Full DHW DGP under both innovation mechanisms. For each of $100$ independent replications, we simulate the process to a common forecast origin
\[
t^\star
=
40+60+300
=
400,
\]
and retain the true filtered state and the true smoothing parameters at that origin. Let $\mu^{\mathrm{plug}}_{t^\star+h\mid t^\star}$ denote the plug-in forecast obtained by iterating the forecast recursion without simulating future innovations. From the same state, we then generate $B_{\mathrm{MC}}=2000$ independent future continuations. Since the one-dimensional Fr\'echet mean under squared Wasserstein loss is obtained by averaging quantile functions, its Monte Carlo approximation is
\[
\qf{\mu^{\mathrm{MC}}_{t^\star+h\mid t^\star}}(u)
=
\frac{1}{B_{\mathrm{MC}}}
\sum_{b=1}^{B_{\mathrm{MC}}}
\qf{\nu^{(b)}_{t^\star+h}}(u).
\]
For $h\in\{1,2,4\}$, we record
\[
D_h
=
W_2^2\left(
\mu^{\mathrm{plug}}_{t^\star+h\mid t^\star},
\mu^{\mathrm{MC}}_{t^\star+h\mid t^\star}
\right).
\]

\begin{figure}[t]
\centering
\includegraphics[width=0.68\textwidth]{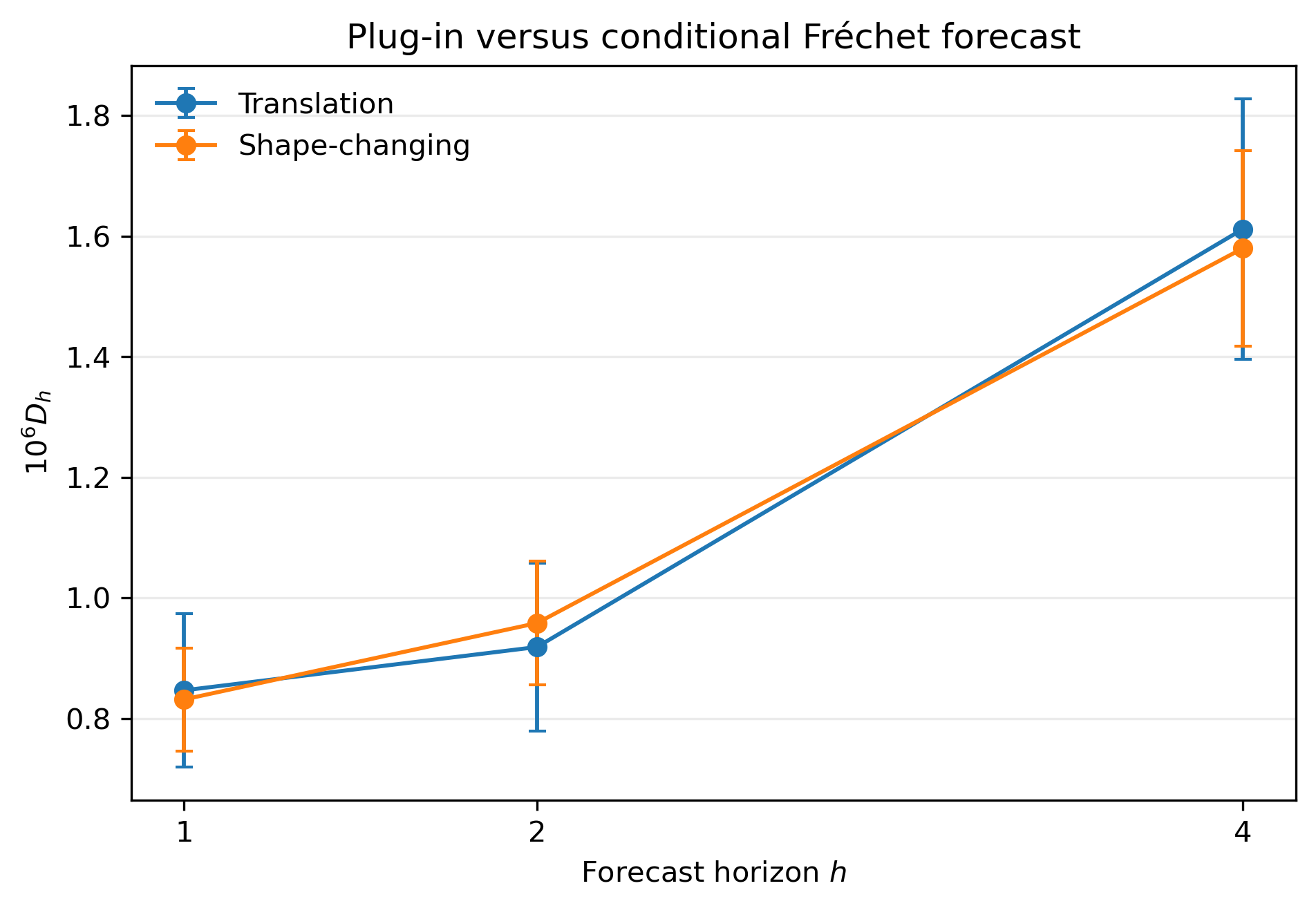}
\caption{Discrepancy between the plug-in multi-step forecast and a nested Monte Carlo approximation to the conditional Fr\'echet mean. The vertical axis reports $10^6D_h$. Points are averages over $100$ independent forecast origins and error bars are Monte Carlo standard errors. The positive value at $h=1$ reflects finite-$B_{\mathrm{MC}}$ approximation error because the population discrepancy is zero at one step.}
\label{fig:appendix-multistep}
\end{figure}

Figure~\ref{fig:appendix-multistep} reports $10^6D_h$. At $h=1$, the population discrepancy is exactly zero, so the observed means of $0.85\times10^{-6}$ under translation innovations and $0.83\times10^{-6}$ under shape-changing innovations provide an empirical noise floor for the nested Monte Carlo approximation. The mean discrepancy at $h=2$ remains close to this level, at $0.92\times10^{-6}$ and $0.96\times10^{-6}$, respectively. At $h=4$ it rises modestly to $1.61\times10^{-6}$ and $1.58\times10^{-6}$. Even at the longest horizon considered, these values are roughly three orders of magnitude smaller than the one-step oracle MSWE in Section~\ref{sec:simulation-study}. Thus the simple plug-in recursion is numerically very close to the conditional Fr\'echet forecast for the horizons and DGP specifications considered here.

\paragraph{Grid sensitivity.}
We assess sensitivity to the numerical representation using the Full DHW DGP with shape-changing innovations, so that all structural components and the nonlinear innovation mechanism are active. Using $20$ paired Monte Carlo replications, we compare
\[
(K,M_x)
\in
\{(99,2001),(199,4001),(399,8001)\}.
\]
The middle grid, $(K,M_x)=(199,4001)$, is used in Section~\ref{sec:simulation-study}. Table~\ref{tab:appendix-grid-sensitivity} compares this grid directly with the finer grid using relative differences. Mean relative differences in the smoothing-parameter estimates are all below $0.05\%$; the corresponding mean relative differences in MSWE and FOE are $0.274\%$ and $0.469\%$. Optimisation succeeds in every paired run on both grids, and none of the paired paths leaves the stored map-grid range. These results indicate that the grid used in the simulation study is sufficiently fine for the reported results.

\begin{table}[t]
\centering
\small
\caption{Sensitivity of the simulation results to the numerical probability and map grids. Relative differences compare $(K,M_x)=(199,4001)$ with $(K,M_x)=(399,8001)$.}
\label{tab:appendix-grid-sensitivity}
\begin{tabular}{lrr}
\toprule
Quantity & Mean relative difference (\%) & Maximum relative difference (\%) \\
\midrule
$\widehat\alpha$ & 0.032 & 0.125 \\
$\widehat\beta$ & 0.042 & 0.133 \\
$\widehat\gamma$ & 0.022 & 0.111 \\
$\widehat\delta$ & 0.026 & 0.071 \\
Training loss & 0.207 & 0.467 \\
MSWE & 0.274 & 1.290 \\
Oracle MSWE & 0.273 & 1.267 \\
FOE & 0.469 & 1.641 \\
\midrule
\multicolumn{3}{l}{Optimisation success rate: $(199,4001)$ grid, 1.000; $(399,8001)$ grid, 1.000.} \\
\multicolumn{3}{l}{Maximum fraction outside the value grid: $(199,4001)$ grid, 0; $(399,8001)$ grid, 0.} \\
\bottomrule
\end{tabular}
\end{table}

\paragraph{Optimisation and tail diagnostics.}
Across the full one-step experiment, the selected solution has a successful optimiser status in $11979$ of the $12000$ fitted model--replication combinations. The $21$ non-success statuses occur only for WES-T under the Full DHW DGP and for DHW-FS under the Level + trend or Full DHW DGPs. In each case the returned candidate lies at the active boundary $\beta=1$ and the optimiser reports ``inequality constraints incompatible''. All correctly specified fits and all fits of the unrestricted DHW model have successful optimiser status. During DHW optimisation, occasional trial parameter values can produce intermediate maps that fail the numerical map checks; these trial points are assigned a large objective value and excluded from the selected solution. This does not prevent convergence of the unrestricted DHW fits.

The stored map grid is used for interpolation rather than truncation: outside $[-100,100]$, the displacement is continued with unit slope. Among the trend-containing DGPs, between $2.2\%$ and $5.8\%$ of Monte Carlo replications contain at least one path ordinate outside the stored grid. Such excursions are therefore uncommon, and the correctly specified FOE is essentially unchanged when they occur. For example, under the Full DHW DGP the mean FOE is approximately $9\times10^{-6}$ with or without an excursion under shape-changing innovations, and is slightly smaller in the small subset of translation replications that leave the stored grid. The forecasting conclusions are therefore not driven by tail continuation.

\paragraph{Complete one-step forecasting results and risk decomposition.}
Table~\ref{tab:appendix-full-forecast} reports the complete one-step MSWE and FOE results with Monte Carlo standard errors. Both the Monte Carlo means and their standard errors are multiplied by $10^3$, matching the scale used in Table~\ref{tab:simulation-forecast}.

\begin{table}[t]
\centering
\small
\caption{Complete one-step-ahead forecasting results. Both the Monte Carlo means and the corresponding Monte Carlo standard errors in parentheses are multiplied by $10^3$.}
\label{tab:appendix-full-forecast}
\begin{tabular}{llcc}
\toprule
DGP & Model & $10^3\times$ MSWE & $10^3\times$ FOE \\
\midrule
\multicolumn{4}{l}{\textit{Panel A: Translation innovations}} \\
Level & WES & 1.6035 (0.0101) & 0.0061 (0.0004) \\
 & WES-T & 1.6068 (0.0102) & 0.0099 (0.0007) \\
 & DHW-FS & 1.7362 (0.0115) & 0.1360 (0.0045) \\
 & DHW & 1.6287 (0.0103) & 0.0308 (0.0010) \\
Level + trend & WES & 17.0956 (0.7670) & 15.4528 (0.7521) \\
 & WES-T & 1.6218 (0.0113) & 0.0157 (0.0015) \\
 & DHW-FS & 11.2159 (0.2254) & 9.6365 (0.2215) \\
 & DHW & 1.6859 (0.0119) & 0.0760 (0.0020) \\
Full DHW & WES & 109.8466 (3.0277) & 108.2432 (2.9885) \\
 & WES-T & 53.2389 (1.7002) & 51.6666 (1.6692) \\
 & DHW-FS & 49.7709 (1.5777) & 48.1475 (1.5539) \\
 & DHW & 1.6260 (0.0106) & 0.0183 (0.0008) \\
\midrule
\multicolumn{4}{l}{\textit{Panel B: Shape-changing innovations}} \\
Level & WES & 1.3737 (0.0075) & 0.0028 (0.0002) \\
 & WES-T & 1.3751 (0.0075) & 0.0039 (0.0002) \\
 & DHW-FS & 1.4857 (0.0086) & 0.1136 (0.0030) \\
 & DHW & 1.3923 (0.0076) & 0.0214 (0.0005) \\
Level + trend & WES & 16.8326 (0.8701) & 15.2188 (0.8579) \\
 & WES-T & 1.6380 (0.0173) & 0.0060 (0.0003) \\
 & DHW-FS & 10.8863 (0.2248) & 9.2573 (0.2205) \\
 & DHW & 1.7023 (0.0179) & 0.0721 (0.0013) \\
Full DHW & WES & 126.9698 (3.4494) & 125.4780 (3.4198) \\
 & WES-T & 62.9839 (1.8289) & 61.4105 (1.8066) \\
 & DHW-FS & 57.0947 (1.6311) & 55.5121 (1.6138) \\
 & DHW & 1.6566 (0.0188) & 0.0092 (0.0004) \\
\bottomrule
\end{tabular}
\end{table}

As a separate check of the theoretical risk decomposition in Section~\ref{sec:theory}, we compute the realised risk-decomposition gap
\[
\operatorname{MSWE}
-
\operatorname{MSWE}_{\mathrm{oracle}}
-
\operatorname{FOE}
\]
within each replication and average this quantity over replications. The population identity concerns the relevant conditional expectation, so the realised gap need not vanish in any individual replication. Across all $24$ model--DGP--innovation combinations, however, the absolute mean gap is never more than $1.67$ Monte Carlo standard errors from zero. We therefore find no evidence of a systematic discrepancy from the theoretical decomposition.


\section{\texttt{dhwsmooth}: Python Package for Distributional Holt--Winters}
\label{app:dhwsmooth}

The accompanying Python package \texttt{dhwsmooth} provides an implementation of the distributional Holt--Winters methodology developed in this paper. The package is available at
\[
\text{\url{https://github.com/wilson-ye-chen/dhwsmooth}}.
\]
It contains routines for initialisation, filtering, parameter estimation, one- and multi-step forecasting, simulation from the DHW recursion, numerical evaluation of monotone transport maps, and visualisation of distribution-valued sample paths and the associated trend and seasonal maps.

The package can be installed directly from the repository using
\begin{verbatim}
pip install git+https://github.com/wilson-ye-chen/dhwsmooth
\end{verbatim}
The following code illustrates the basic workflow for fitting a DHW model to a distribution-valued time series. Suppose that the rows of \texttt{y} contain discretised quantile functions, with the first \texttt{n\_init} observations reserved for initialisation:
\begin{verbatim}
import numpy as np
from dhwsmooth.core import DistHoltWinters

m = 4
x_grid = np.linspace(-100, 100, 4001)

model = DistHoltWinters(m=m, x_grid=x_grid)

y_init = y[:n_init]
y_fit = y[n_init:]

init = model.initialise(y_init)

result = model.fit(
    y_fit,
    init,
    start=n_burn,
)

mu_hat, state, history = model.predict(
    y_fit,
    init,
    return_state=True,
    return_history=True,
)

forecast = model.forecast(
    state,
    h=4,
)
\end{verbatim}
Here \texttt{mu\_hat} contains the fitted one-step-ahead forecast distributions, while \texttt{state} contains the final filtered level, trend, seasonal maps, and seasonal phase. Setting \texttt{return\_history=True} additionally records the evolution of these filtered components through time.

The package also provides the class \texttt{DistHoltWintersSampler} for simulating directly from the DHW recursion. The transport-map utilities in \texttt{dhwsmooth.maps} implement the one-dimensional interpolation, inversion, and Wasserstein-loss calculations used by the model. Plotting utilities in \texttt{dhwsmooth.plot} include density-path visualisations and functions for displaying the evolution of the trend and seasonal transport maps. For example,
\begin{verbatim}
from dhwsmooth.plot import (
    plot_density_path,
    plot_map_history,
    plot_seasonal_cycle,
    plot_seasonal_history,
)
\end{verbatim}
provides visualisation tools for distribution-valued paths and the filtered DHW components.


\section{Additional Empirical Details}
\label{app:empirical}

\subsection{Numerical representation and estimation}
\label{app:empirical-numerics}

Each distribution is represented by $K=101$ linearly interpolated sample quantiles at $u_k=(k-1/2)/K$. To make the numerical quantile representation strictly increasing, raw ordinates $q^{\mathrm{raw}}_{t,k}$ are adjusted recursively as
\[
q_{t,1}=q^{\mathrm{raw}}_{t,1},\qquad
q_{t,k}=\max\{q^{\mathrm{raw}}_{t,k},q_{t,k-1}+\varepsilon d_k\},
\quad k=2,\ldots,K.
\]
For the hourly applications, $d_k=u_k-u_{k-1}$ and $\varepsilon=10^{-6}\max(s,1)$; for retail, $d_k=1$ and $\varepsilon=10^{-10}\max(s,1)$. Here $s$ is the standard deviation of the raw quantile ordinates over the complete study window, including the test dates. All methods use the same adjusted observations for fitting and scoring.

Value-map knots are obtained from 201 equally spaced quantiles of the pooled adjusted ordinates dated before the test period. Duplicate hourly knots are removed, giving 199 knots for NYC and 201 for electricity. Retail retains 201 knots with a minimal positive spacing. Map displacements are interpolated linearly, with constant boundary displacement outside the stored range; inverse evaluation swaps the input and output knots. The maps are not recentred during filtering.

NYC stations are identified by the source data's station-complex codes. The hourly calendars retain unavailable clock-change positions. NYC excludes 02:00 on the spring clock-change dates and 01:00 on the autumn dates. Electricity first aggregates the quarter-hour timestamps $h+15$, $h+30$, $h+45$ and $h+60$ minutes into the hour labelled $h$, dividing their sum by four, and excludes the hour labelled 01:00 on each spring and autumn clock-change date. At unavailable dates, DHW advances the level through the trend map, retains the trend and seasonal maps and advances the seasonal phase; that date contributes no fitting or test loss.

\paragraph{Initialisation and chronology.}
Initial states follow Section~\ref{sec:dhw-initialisation}. Within the initial sample, quantiles are averaged at each seasonal position; their barycentre defines the initial level, the trend map is the identity and each initial seasonal map transports that level to its corresponding seasonal average. The sample comprises complete cycles before the first unavailable date, or all complete cycles within the initial block when there are no gaps:
\begin{center}
\begin{tabular}{@{}llll@{}}
\toprule
Application & Initial sample & Cycles & Subsequent unscored warm-up \\
\midrule
NYC & 13 Jan.--9 Mar. 2017 & 8 weeks & 10 Mar.--31 Dec. 2017 \\
Electricity & 2 Jan.--18 Mar. 2012 & 11 weeks & 19 Mar.--31 Dec. 2012 \\
Retail & Apr. 1982--Mar. 1993 & 11 years & Apr.--Dec. 1993 \\
\bottomrule
\end{tabular}
\end{center}
These samples contain 1,344, 1,848 and 132 distributions, respectively. The warm-up contains 7,128 calendar hours (7,126 observed) for NYC, 6,912 hours (6,910 observed) for electricity and nine months for retail. Each parameter candidate is filtered through this warm-up before fitting losses are accumulated. Estimation uses 2018 for NYC, 2013 for electricity and 1994--2005 for retail, containing 8,758, 8,758 and 144 observed targets. The resulting state is carried into the test period without reinitialisation. Parameters remain fixed, and each one-step forecast precedes the observation update.

\paragraph{Parameter estimation.}
The shared \texttt{dhwsmooth} fitter uses SLSQP in $(\alpha,\delta,\gamma)$, where $\delta=\alpha\beta$, with
\[
10^{-6}\leq\alpha\leq1,\qquad 0\leq\delta\leq\alpha,\qquad 0\leq\gamma\leq1,
\qquad \beta=\delta/\alpha.
\]
For each hourly application, a preliminary fit from $(0.9,0,0.01)$ minimises the fitting loss divided by $10^4$; its solution initialises the final fit on the original loss scale. Retail uses the starting point $(0.3,0.1,0.3)$ directly. All three final fits terminate successfully; Table~\ref{tab:real-parameters} reports the estimates. The hourly loss evaluation preserves the complete calendar while scoring only observed fitting targets.

\begin{table}[tbp]
\centering
\caption{Fitted DHW smoothing parameters. The optimiser uses $\delta=\alpha\beta$.}
\label{tab:real-parameters}
\small
\begin{tabular}{@{}lrrrr@{}}
\toprule
Application & $\widehat\alpha$ & $\widehat\delta$ & $\widehat\beta$ & $\widehat\gamma$ \\
\midrule
NYC & 1.000000 & $1.35\times10^{-11}$ & $1.35\times10^{-11}$ & 0.006525 \\
Electricity & 0.895891 & 0.000448 & 0.000500 & 0.032499 \\
Retail & 0.327512 & 0.002716 & 0.008292 & 0.263023 \\
\bottomrule
\end{tabular}
\end{table}

\subsection{Benchmark specifications}
\label{app:empirical-benchmarks}

Let $\mathcal I$ contain the observed fitting dates and $\mathcal I_r\subseteq\mathcal I$ those dates whose preceding $r$ calendar observations are also available. Write $\langle f,g\rangle_K=K^{-1}\sum_{k=1}^Kf(u_k)g(u_k)$ and $\|f\|_K^2=\langle f,f\rangle_K$. The four distributional benchmarks are nonseasonal specifications with fixed lag orders. Their parameters, reference means and fitted maps are estimated once and held fixed during testing. Lagged observations are updated after each forecast; an unavailable observation is replaced by that benchmark's own forecast. Seasonal naive uses the latest available observation at the same seasonal position, while seasonal mean averages all previously observed quantile functions at that position.

\paragraph{Consecutive-transport ATM(1).}
We use $\mathrm{ATM}_d(1)$ of \citet{atmzhu2023}. For a strictly increasing continuous bijection $T$ of an interval onto itself, define
\[
[aT](x)=
\begin{cases}
(1-a)x+aT(x), & 0\leq a\leq1,\\
(1+a)x-aT^{-1}(x), & -1\leq a<0.
\end{cases}
\]
With $T_{t-1}=\otmap{\nu_{t-2}}{\nu_{t-1}}$, the forecast quantile function is $[\widehat aT_{t-1}]\circ\qf{\nu_{t-1}}$. We estimate $a\in[-1,1]$ by minimising midpoint-quantile prediction loss over $\mathcal I_2$, considering both sign branches. This predictive criterion differs from the original paper's Lebesgue-integrated map-regression criterion. Transport interpolation and tail continuation follow the DHW convention.

\paragraph{WAR(1).}
For the scalar model of \citet{zhang2022wasserstein}, let $\bar\nu$ be the fitting-block Wasserstein barycentre. We use
\[
\widehat b=
\frac{\displaystyle\sum_{t\in\mathcal I_1}
\left\langle\qf{\nu_{t-1}}-\qf{\bar\nu},\qf{\nu_t}-\qf{\bar\nu}\right\rangle_K}
{\displaystyle\sum_{t\in\mathcal I}
\left\|\qf{\nu_t}-\qf{\bar\nu}\right\|_K^2},
\]
setting $\widehat b=0$ for a zero denominator. Forecast ordinates are $\widehat b\qf{\nu_{t-1}}+(1-\widehat b)\qf{\bar\nu}$, with increasing rearrangement if they cross. The coefficient is not clipped.

\paragraph{DAM--UQ.}
For the uniform-quantile model of \citet{ghodrati2024distributional}, a fixed support $[L,H]$ extends the pre-test quantile range by $10\%$ of that range at each end. Define $\widetilde Q_t=(\qf{\nu_t}-L)/(H-L)$, append endpoint pairs $(0,0)$ and $(1,1)$ and interpolate linearly. For each coefficient $a$, the profile map is
\[
\widehat A_a=|\mathcal I_1|^{-1}
\sum_{t\in\mathcal I_1}\widetilde Q_t\circ[a\widetilde Q_{t-1}]^{-1}.
\]
The coefficient minimises $\sum_{t\in\mathcal I_1}\|\widehat A_a\circ[a\widetilde Q_{t-1}]-\widetilde Q_t\|_K^2$ over $[-0.99,0.99]$, using a 41-point grid followed by bounded refinement around local minima. The profile map is stored on $\{0,u_1,\ldots,u_K,1\}$. Forecasts are transformed back by $L+(H-L)\widehat A_{\widehat a}\circ[\widehat a\widetilde Q_{t-1}]$. The support is fixed throughout testing, and predictors outside it are rejected.

\paragraph{DoDRM.}
We apply \citet{ghodrati2022distribution} to lag-one pairs, fitting a single nondecreasing value map $\widehat T$ to minimise
\[
\sum_{t\in\mathcal I_1}\|T\circ\qf{\nu_{t-1}}-\qf{\nu_t}\|_K^2.
\]
Isotonic regression pools all predictor--response ordinates with weight $1/K$. Prediction uses $\widehat T\circ\qf{\nu_{t-1}}$, linear interpolation between fitted thresholds and constant endpoint values outside the fitted predictor range. Flat segments are retained.

\subsection{Fitted distributional components}
\label{app:empirical-components}

The retail component plots use filtered states after December 2005, December 2006 and December 2018. Each state is computed after incorporating that month's observation, with the smoothing parameters estimated from 1994--2005 held fixed. The trend displacement is evaluated as $\Bmap_t(x)-x$ on the same value range at all three dates.

For the seasonal effects, let $t_\star$ denote December 2005 and define the common reference values
\[
x_\star(u)
=\qf{\eta_{t_\star+1}}(u)
=\Bmap_{t_\star}\!\left\{\qf{\lambda_{t_\star}}(u)\right\},
\qquad u\in\{0.1,0.5,0.9\}.
\]
They are approximately $2.719$, $4.700$ and $6.240$ in natural-log turnover units. At each December state $t$, the displayed January--December cycle is
\[
d_{t,h}(u)=\Smap_{j(t,h)}\{x_\star(u)\}-x_\star(u),
\qquad h=1,\ldots,12,
\]
where $j(t,h)$ is defined in \eqref{eq:dhw-seasonal-forecast-index}. Thus each curve evaluates the twelve stored seasonal maps at a fixed reference quantile. Holding these reference values constant across months and state dates makes differences attributable to the maps. A positive displacement raises the corresponding log-turnover value, and unequal displacements across quantiles indicate changes in distributional shape. The displacements $d_{t,h}(u)$ need not be ordered in $u$. Monotonicity requires that, for $u_1<u_2$,
\[
x_\star(u_1)+d_{t,h}(u_1)<x_\star(u_2)+d_{t,h}(u_2),
\]
which holds because each seasonal map is increasing. The displacement curves can therefore cross while the transformed quantiles remain ordered. All displayed map evaluations lie inside the stored value grid.

\begin{figure}[p]
\centering
\includegraphics[width=\linewidth,height=0.74\textheight,keepaspectratio]{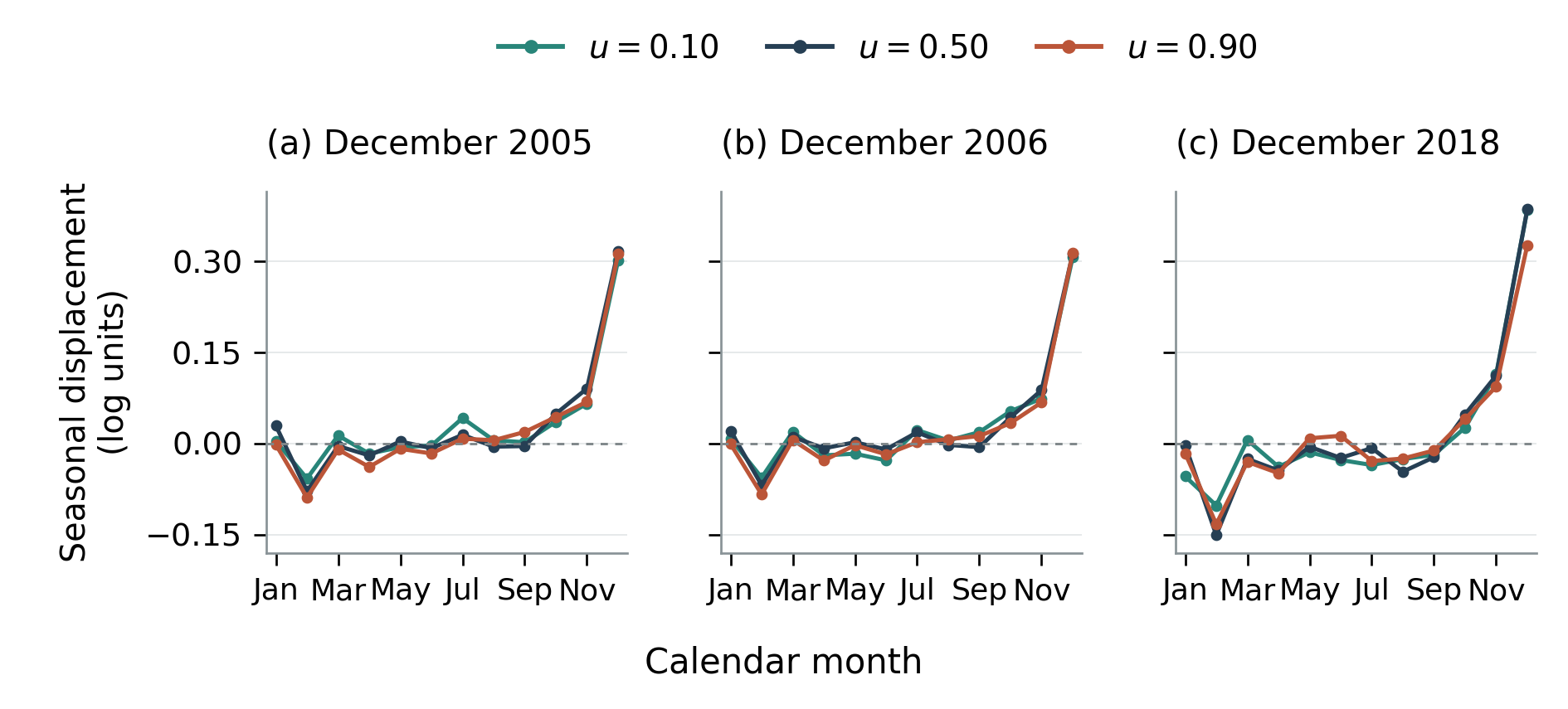}
\caption{Complete fitted seasonal cycles for Australian retail after December 2005, December 2006 and December 2018. Each panel evaluates the twelve seasonal maps at the same reference values $x_\star(0.1)$, $x_\star(0.5)$ and $x_\star(0.9)$ defined in Appendix~\ref{app:empirical-components}. Curves therefore compare map displacements in natural-log turnover units on a common scale. The dashed zero line denotes an identity effect.}
\label{fig:retail-seasonal-states}
\end{figure}

\subsection{Uncertainty in loss comparisons}
\label{app:empirical-uncertainty}

Paired circular block-bootstrap resampling uses 19,999 draws, with blocks of 168 calendar hours for NYC and electricity and 12 months for retail. All methods share each draw's dates, including unavailable positions, and losses are averaged over observed targets. Pointwise 95\% intervals are the percentile intervals of the resampled percentage loss reductions. Two-sided tests centre the resampled mean loss differences at their observed values; Holm adjustment covers the twelve DHW comparisons with the four distributional benchmarks. Inference is conditional on the fitted models and the stations, clients and retail series included in the analysis. Resampling paired calendar blocks preserves dependence within each block and assesses uncertainty in average forecast-loss differences over time.

To assess sensitivity to dependence across seasonal cycles, we repeat the calculation with blocks of 336 and 672 hours for NYC and electricity, and 24 and 48 months for retail. All twelve percentage-loss-reduction intervals remain above zero, and all Holm-adjusted bootstrap $p$-values remain below 0.001. With four-cycle blocks, the smallest lower endpoints across the four distributional comparisons are 93.23\% for NYC, 60.63\% for electricity and 89.98\% for retail.

\begin{figure}[p]
\centering
\includegraphics[width=\linewidth,height=0.76\textheight,keepaspectratio]{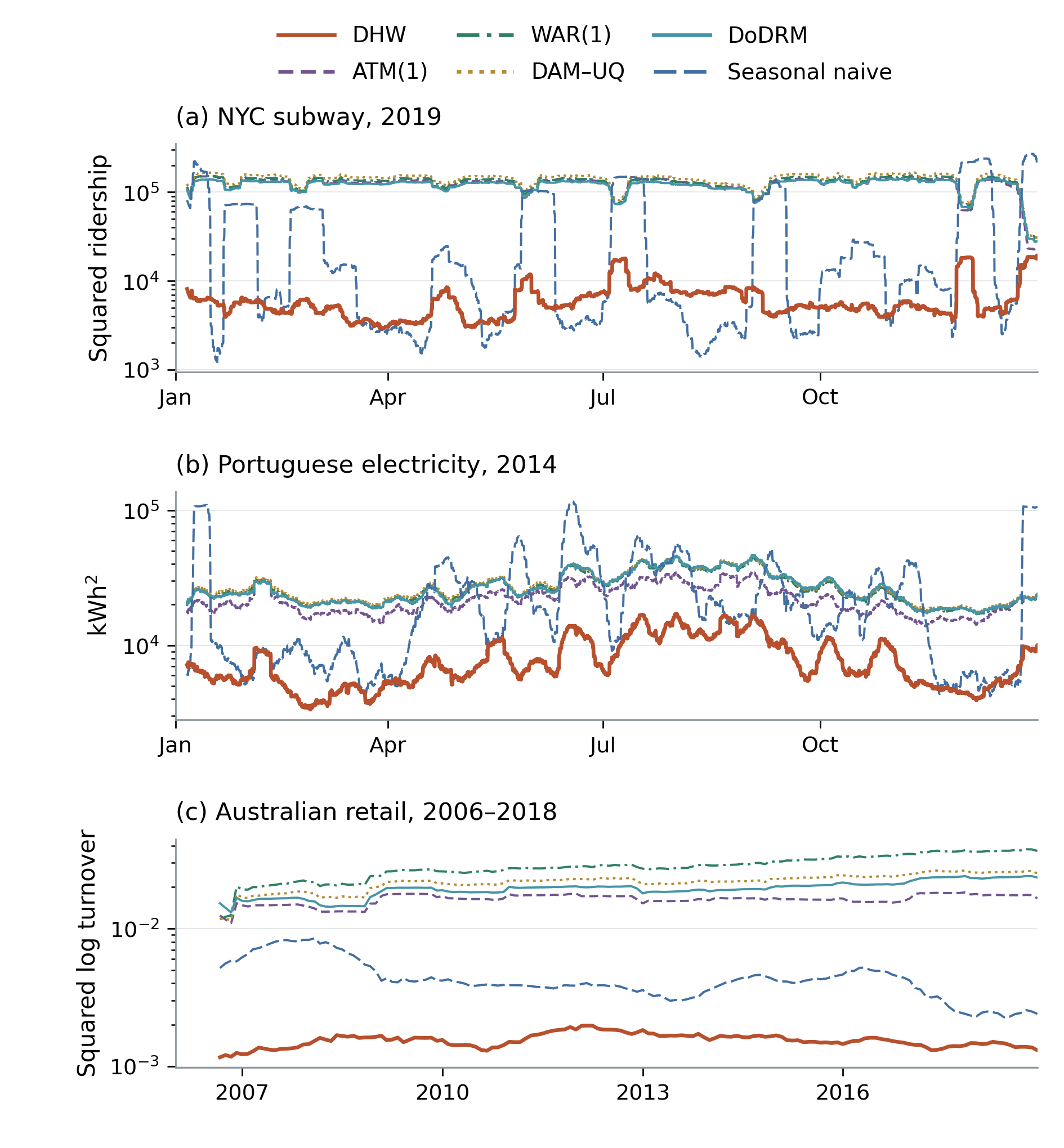}
\caption{Rolling mean squared Wasserstein losses for DHW, the four distributional benchmarks and seasonal naive. Windows contain 168 calendar hours or 12 months and require at least 70\% observed scores. All vertical axes are logarithmic. Retail losses are in squared natural-log turnover units; absolute loss levels are not comparable across applications.}
\label{fig:real-losses}
\end{figure}

The lower mean loss of DHW is also present in each of the twelve calendar months of both hourly test periods and each of the thirteen retail test years. Figure~\ref{fig:real-losses} shows the loss paths averaged over one seasonal cycle. Loss differences vary over time, so the approximate bootstrap intervals are interpreted together with this consistency across subperiods.

\FloatBarrier

\end{document}